# Adaptive tempered reversible jump algorithm for Bayesian curve fitting

Zhiyao Tian[1,2,3], Anthony Lee[3], Shunhua Zhou[1,2*],

[1] *Key Laboratory of Road and Traffic Engineering of the Ministry of Education, Tongji University, Shanghai, China*

[2] *Shanghai Key Laboratory of Rail Infrastructure Durability and System Safety, Tongji University, Shanghai, China*

[3] *School of Mathematics, University of Bristol, Bristol, United Kingdom*

**Corresponding author, E-mail address: zhoushh@tongji.edu.cn.* (*Shunhua Zhou*)

**Abstract:** Bayesian curve fitting plays an important role in inverse problems, and is often addressed using the Reversible Jump Markov Chain Monte Carlo (RJMCMC) algorithm. However, this algorithm can be computationally inefficient without appropriately tuned proposals. As a remedy, we present an adaptive RJMCMC algorithm for the curve fitting problems by extending the Adaptive Metropolis sampler from a fixed-dimensional to a trans-dimensional case. In this presented algorithm, both the size and orientation of the proposal function can be automatically adjusted in the sampling process. Specifically, the curve fitting setting allows for the approximation of the posterior covariance of the *a priori* unknown function on a representative grid of points. This approximation facilitates the definition of efficient proposals. In addition, we introduce an auxiliary-tempered version of this algorithm via non-reversible parallel tempering. To evaluate the algorithms, we conduct numerical tests involving a series of controlled experiments. The results demonstrate that the adaptive algorithms exhibit significantly higher efficiency compared to the conventional ones. Even in cases where the posterior distribution is highly complex, leading to ineffective convergence in the auxiliary-tempered conventional RJMCMC, the proposed auxiliary-tempered adaptive RJMCMC performs satisfactorily. Furthermore, we present a realistic inverse example to test the algorithms. The successful application of the adaptive algorithm distinguishes it again from the conventional one that fails to converge effectively even after millions of iterations.

## 1 Introduction

This paper explores the inference of a continuous function, $f$, with respect to space or time, $x$, given a set of observed data, where $f$ is a piecewise polynomial function [1]:

$$f(x)=\sum_{i=1}^{n}a_i b_i(x), \tag{1}$$

where, $n$ represents the number of knots; $\mathbf{a}=(a_1,\ldots,a_n)$ are the unknown coefficients; $b_i(x)$ are basis functions. For instance, $b_i(x)= \max\{x - r_i, 0\}$ for the piecewise linear case; $\mathbf{r}=(r_1,\ldots,r_n)$ denotes the locations of the knots. In this context, the inference of $f$ involves inferring the joint parameters $\mathbf{m}$, where $\mathbf{m} = (n,\mathbf{r},\mathbf{a})$.

A straightforward approach to infer $\mathbf{m}$ is within a deterministic optimization framework. However, in certain situations, a unique solution may not be guaranteed [2]. Alternatively, the Bayesian approach entails making inferences from a posterior distribution defined in the model space. One advantage of this approach is its ability to account for model uncertainty beyond single model selection [3]. This aspect is particularly pivotal in cases involving complex models and limited data. In the context of this paper, a Bayesian approach facilitates the generation of diverse, approximate samples of $f$ from the posterior distribution, thus providing a quantifiable interpretation of uncertainty. Nevertheless, a limitation arises from its relatively high computational demands [4]. Specifically, Bayesian inference of $\mathbf{m}$, where the number of parameters itself is unknown *a priori*, is typically addressed using the trans-dimensional sampler – the Reversible Jump Markov Chain Monte Carlo (RJMCMC) algorithm [5]. This algorithm, while effective, can be computationally intensive in certain scenarios.

It is convenient, though not rigorous, to classify curve fitting problems into two categories: regression problems and inverse problems, based on the relation between $f$ and the observed data. Let $\mathbf{D}$ be the observed dataset that consists of $k$ observed responses $d_i$ at known locations $x_i$, i.e., $\mathbf{D} =\{(x_i, d_i)\}$, $i$=1,…,$k$. In regression problems, $f$ is directly informed by the observed data, i.e., $d_i = f(x_i)+e_i$, where $e_i$ are random errors; while in inverse problems, $f$ is indirectly informed by the data via a mapping function $g$ (also called the forward model), i.e., $d_i = g(f, x_i)+e_i$. Essentially, the regression problem is a special case of an inverse problem with $g(f, x_i) = f(x_i)$. However, the analytical and explicit relation between $\mathbf{D}$ and $f$ in regression problems allows for a relatively straightforward design of RJMCMC with low computational complexity. For instance, Denison et al. [6] designed an algorithm that samples *a posterior* for $n$ and $\mathbf{r}$ assuming the coefficient $\mathbf{a}$ is given by its corresponding least squares estimate. DiMatteo et al. [7] set conjugate priors for $\mathbf{a}$, and integrated the coefficients out of the posterior analytically, leaving $n$ and $\mathbf{r}$ as the unknowns to be sampled. Similar work can be seen in Fan et al. [8], Poon and Wang [9], and Chen and Chen. [10]. However, for most inverse problems, conjugate priors for the coefficients $\mathbf{a}$ are not available due to the nature of the forward model $g$. In some practical problems, like the recovery of continuous physical properties of the earth [11–12] or the inversion of pressure field on engineering structures [13], the forward model can be non-linear or implicit. In such cases, the parameters $\mathbf{m}=(n,\mathbf{r},\mathbf{a})$ have to be sampled jointly, potentially imposing a significant computational burden on the samplers.

Enhancing the efficiency of RJMCMC is significant for the aforementioned inverse problems, given the computational complexity of the sampling process as well as the computational cost of the forward model itself. Numerous methods have been proposed to address this challenge. Parallel Tempering (PT) [14], an auxiliary method to accelerate RJMCMC, involves running multiple Markov chains in parallel with successively relaxed (tempered) densities [15–16]. Tempered chains allow for more exploration in the parameter space, and swaps between the chains enable sample transfers. With assistance of these auxiliary chains, the un-tempered chain can frequently visit high probability regions while maintaining detailed balance. Another aspect involves enhancing the efficiency of within-chain sampling, specifically the proposal function, which is known to be crucial for the convergence of the Markov chain. For example, Green and Mira [17] introduced a delayed rejection proposal scheme to improve the acceptance rate of RJMCMC. Dosso et al. [18] developed a principal-component space proposal scheme and demonstrated its efficacy in geoacoustic profile estimation. Farr et al. [19] proposed efficient trans-dimensional proposals by approximating the posterior probability density from a fixed-dimensional MCMC run. Gallagher et al. [20] and Zhang [21] scaled the proposal function dynamically according to the acceptance rate. In general, many creative and effective ideas have been presented to construct an efficient proposal for RJMCMC.

A widely adopted and successful approach in fixed-dimensional settings is the Adaptive Metropolis (AM) algorithm proposed by Haario et al. [22]. AM utilizes an approximation of the posterior covariance in a random walk Metropolis algorithm. However, extending this approach to the trans-dimensional setting is not straightforward, because the coefficients in different models are not always related to each other [23]. Nevertheless, in the context of curve fitting, each model defines an approximation of the unknown curve, and the posterior covariance of the curve's values at different points can be well-defined. Specifically, we will demonstrate how to use an interpolation technique, inherent to curve fitting problems, to define adaptive proposals for both the inter- and intra-model proposals by collecting sampling history in different models. Additionally, we will illustrate the complementary and straightforward use of the auxiliary PT algorithm to enhance the robustness of the adaptation.

The structure of this paper is as follows: Section 2 provides a brief introduction to the curve fitting problem and conventional RJMCMC. Section 3 introduces the adaptive RJMCMC algorithm in detail. Section 4 presents numerical examples with a series of controlled experiments to examine the efficiency of the algorithms. Finally, in Section 5, a practical application for the inversion of the pressure field on an engineering structure is presented.

## 2 The Bayesian curve fitting and conventional RJMCMC

### 2.1 *The curve fitting problem*

For clarity, Eq. (1) is reformulated into an interpolation form, essentially equivalent to the following expression:

$$f(x) = \mathbf{I}_{\mathbf{r}}(x)\mathbf{a}, \tag{2}$$

where, as visually presented in Fig. 1, $\mathbf{a}=(a_1,\ldots,a_n)$ represents unknown interpolating values on the $n$ interpolation knots; $\mathbf{I}_{\mathbf{r}}(x)$ is the interpolating matrix that contains parameters denoting locations of the knots $\mathbf{r}=(r_1,\ldots,r_n)$. The derivation of $\mathbf{I}_{\mathbf{r}}(x)$ for typical piecewise constant, linear, and spline functions can be found in Press et al. [24]. Again, Eq. (2) is essentially identical to Eq. (1), and the objective is to infer the joint parameters $\mathbf{m}=(n,\mathbf{r},\mathbf{a})$ using the observed dataset $\mathbf{D}$.

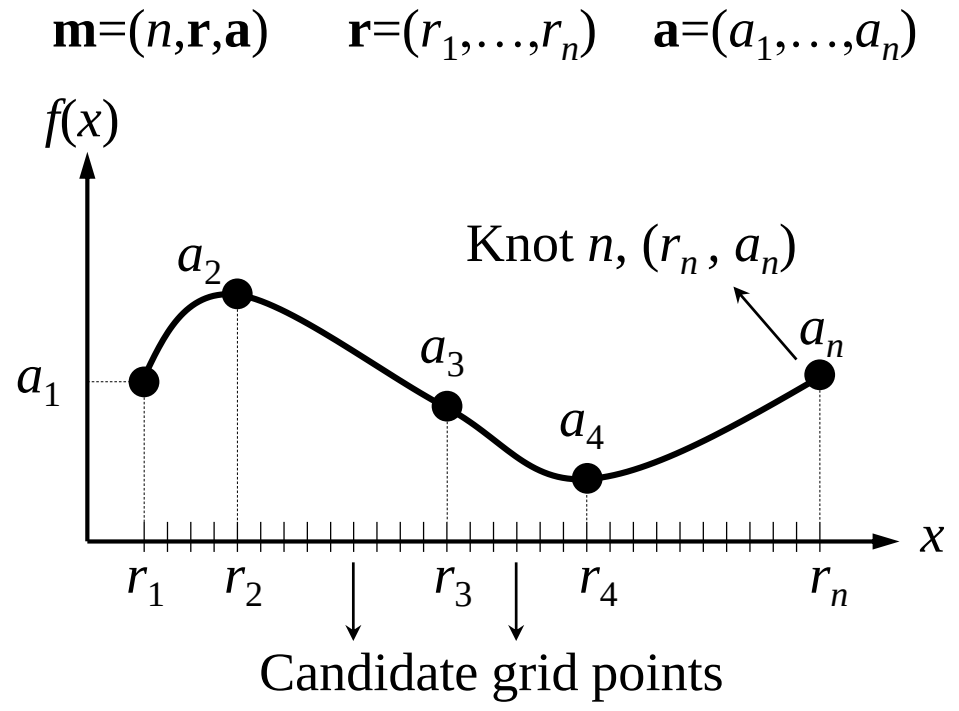

**Fig. 1**. Visual representation of the unknown parameters for the curve fitting problems (an example of piecewise spline).

Generally, $n$ is a discrete random variable taking values in $\mathbf{N}_+$; $a_i$ ($i$=1,…,$n$) is a real-valued, continuous random variable; $r_i$ ($i$=1,…,$n$) can be regarded as either a continuous random variable [7] or a discrete random variable [6]. In this paper, for mathematical convenience, $r_i$ is defined as a discrete random variable taking values in a candidate grid of points (Fig. 1). The grid points are uniformly and densely set within the definition domain of $f$. The curve fitting capacity of the parameters is influenced by the grid density. Thus, to ensure the generality of the curves, it is necessary to set a sufficiently dense grid, meeting the accuracy requirements of a specific problem. Grid density can be determined based on prior experience or through trial testing methods. By gradually increasing the grid density in trial tests, one can monitor the results of inversion. Once the results no longer change with increasing density, the grid density is considered sufficient. While not essential, it is recommended to fix two knots at the endpoints of the domain of $f$ (the ends of the candidate points) to avoid extrapolation of $f$, specifically, fixing $r_1$ and $r_n$ at the ends of the candidate points, with $n\geq 2$.

### 2.2 *The Bayesian approach*

The inference of $\mathbf{m}$ in a Bayesian framework is expressed as follows:

$$p(\mathbf{m}|\mathbf{D}) = \frac{p(\mathbf{D}|\mathbf{m})\,p(\mathbf{m})}{p(\mathbf{D})}, \tag{3}$$

where, $p(\mathbf{m}|\mathbf{D})$ represents the posterior density of parameters $\mathbf{m}$ given observed dataset $\mathbf{D}$; $p(\mathbf{m})$ is the prior density of $\mathbf{m}$; $p(\mathbf{D}|\mathbf{m})$ is the likelihood function; $p(\mathbf{D})$ is a normalizing factor ensuring $p(\mathbf{m}|\mathbf{D})$ integrate to one.

The prior density can be expanded as

$$p(\mathbf{m}) = p(n)p(\mathbf{r}|n)p(\mathbf{a}|n,\mathbf{r})\,. \tag{4}$$

The prior density reflects prior information on the parameters, determined on a case-by-case basis. For example, the priors for $n$ and $a_i$ could be taken as uniform distributions within physically plausible bounds. A simple prior for $\mathbf{r}$ given $n$ is the distribution of $n$ points sampled without replacement from the candidate grid points.

The likelihood function measures the fit between observed data and that predicted with curves $f_{\mathbf{m}}$ determined by model $\mathbf{m}$. In cases where unbiased measurement error is dominant, a zero-mean Gaussian distribution is typically assumed for likelihood, as motivated by the Central Limit Theorem, i.e.,

$$p(\mathbf{D}|\mathbf{m}) = \frac{1}{(2\pi\bar{\sigma}_e^2)^{k/2}} \exp\left(-\frac{\sum_{i=1}^{k}[d_i - g(f_{\mathbf{m}}, x_i)]^2}{2\bar{\sigma}_e^2}\right), \tag{5}$$

where, $g$ is the forward model; $\bar{\sigma}_e$ is the estimated standard deviation of the measurement errors. Although it is beyond the focus of this paper, $\bar{\sigma}_e$ can also be inferred in a Bayesian framework [18]. Since the measurement instruments of data can be generally known in advance in most inverse problems, $\bar{\sigma}_e$ is treated here as a constant estimated according to the precision of measurement instruments.

### *2.3 The conventional RJMCMC*

Typically, obtaining an analytical solution for Eq. (3) is challenging. Hence, it is common to employ RJMCMC to estimate the posterior distribution. RJMCMC draws samples from the posterior distribution by generating candidate parameters $\mathbf{m}^*$ based on the current states $\mathbf{m}_c$ via a proposal function $q(\mathbf{m}^*|\mathbf{m}_c)$; then accepting the new sample with an acceptance ratio $\alpha(\mathbf{m}^*|\mathbf{m}_c)$. To address the transition between models with different dimensions, the proposal function is designed to involve both inter-model ("birth" and "death") and intra-model ("move") steps. Typical proposals of the conventional RJMCMC for the curve fitting problem can be designed as follows:

1) Birth step (Fig. 2a): increase the number of parameters ($n$) by 1; add a new knot $r_b$ randomly and uniformly at a candidate point that was not occupied; and assign a value to this new knot. The value $a_b$ for the new knot is generated based on a Gaussian perturbation to previous interpolated state value $a_p$, following $a_b \sim N(a_p, \sigma_b^2)$, where $\sigma_b^2$ is the variance of the proposed increments.

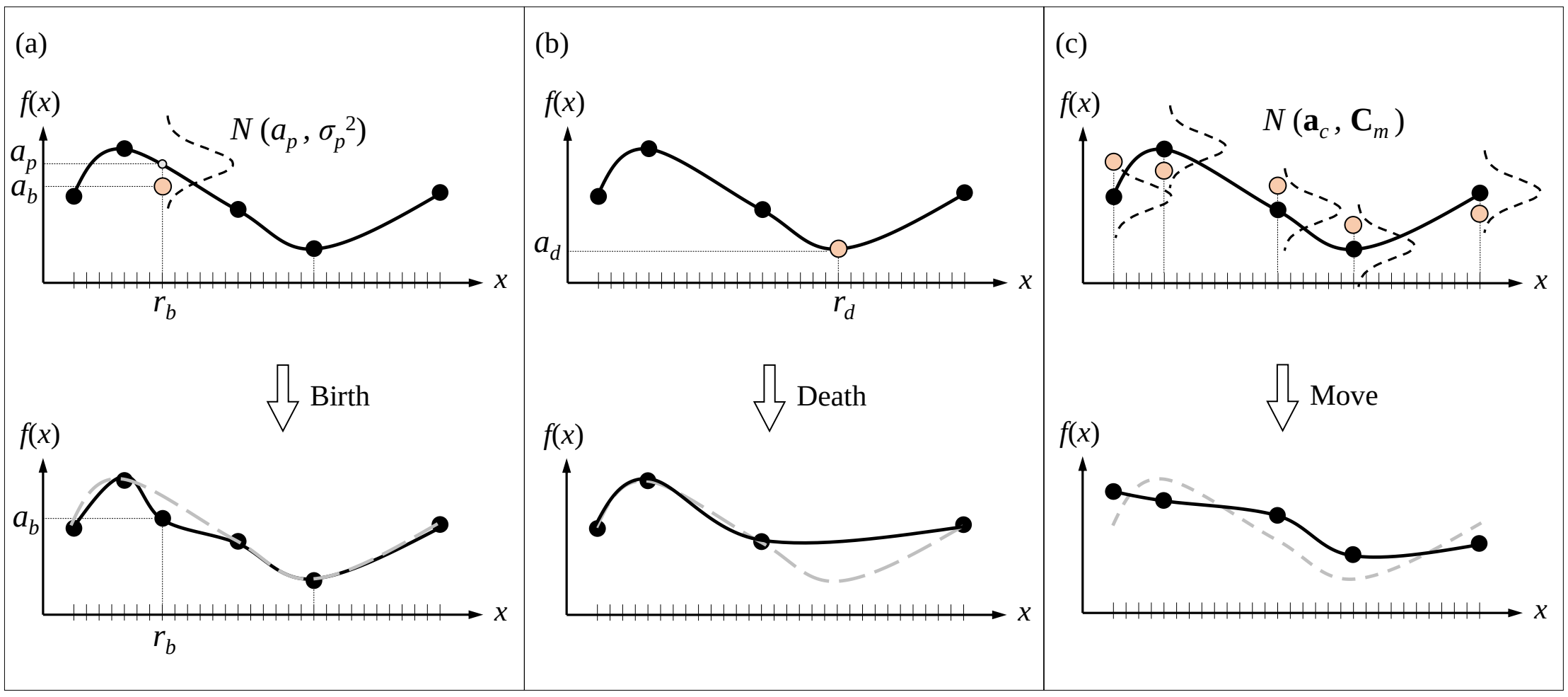

**Fig. 2**. The proposal function: (a) birth; (b) death; (c) move. (an example of spline)

2) Death step (Fig. 2b): decrease the number of parameters ($n$) by 1; and choose one existing knot randomly and uniformly to delete, e.g., delete $r_d$ and the corresponding coefficient $a_d$. It is noteworthy that the knots at the endpoints cannot be selected for deletion to prevent extrapolation of $f$.

3) Move step (Fig. 2c): keep the number ($n$) and location ($\mathbf{r}$) of the existing knots unchanged; and then propose a Gaussian perturbation for the current values $\mathbf{a}_c$, following $N(\mathbf{a}_c, \mathbf{C}_m)$, where $\mathbf{C}_m$ is the covariance matrix for the proposal.

One of the three proposals is applied at each iteration with equal probability. To maintain detailed balance, the new proposal is randomly accepted with an acceptance ratio,

$$\alpha(\mathbf{m}^*|\mathbf{m}_c) = \min\{1, \frac{p(\mathbf{D}|\mathbf{m}^*)p(\mathbf{m}^*)q(\mathbf{m}_c|\mathbf{m}^*)}{p(\mathbf{D}|\mathbf{m}_c)p(\mathbf{m}_c)q(\mathbf{m}^*|\mathbf{m}_c)}|\mathbf{J}|\}, \tag{6}$$

where, $|\mathbf{J}|$ is the Jacobian matrix of the transformation from $\mathbf{m}_c$ to $\mathbf{m}^*$. Derivations of the acceptance ratios for birth, death, and move are provided in Appendix A.

Consequently, through sampling with the proposal function $q(\mathbf{m}^*|\mathbf{m}_c)$ and acceptance ratio $\alpha(\mathbf{m}^*|\mathbf{m}_c)$ on the Markov chain, the samples will eventually converge to the targeted posterior. After convergence of the chains, a set of posterior samples $\{\mathbf{m}_s, s=1,\ldots,S\}$, can be extracted to estimate the posterior distribution:

$$p(\mathbf{m}|\mathbf{d}) \approx \frac{1}{S}\sum_{s=1}^{S}\delta(\mathbf{m}-\mathbf{m}_s). \tag{7}$$

$$p(f|\mathbf{d}) \approx \frac{1}{S}\sum_{s=1}^{S}\delta(f-f_{\mathbf{m}_s}). \tag{8}$$

where, $\delta(\cdot)$ is the Dirac delta function. The expectation of the function reduces to:

$$E(f|\mathbf{d}) = \int p(\mathbf{m}|\mathbf{d}) f_{\mathbf{m}} d\mathbf{m} \approx \frac{1}{S}\sum_{s=1}^{S} f_{\mathbf{m}_s}. \tag{9}$$

## 3 Adaptive proposal RJMCMC

### *3.1 The key idea*

Selecting an appropriate proposal variance/covariance for RJMCMC is crucial yet challenging. Proposal increments that are either "too large or too small" can impede the convergence of the Markov chain. The adaptation of increments in fixed-dimensional MCMC is appealing [22]. In Haario's algorithm [22], a sampling history {$\mathbf{m}_1$, $\mathbf{m}_2$,…, $\mathbf{m}_{t-1}$} informs the proposal covariance at the $t$-th step, denoted as $\mathbf{C}_{pt}$. It is calculated as $\mathbf{C}_{pt} = s_d\mathrm{Cov}(\mathbf{m}_1, \mathbf{m}_2, \ldots, \mathbf{m}_{t-1}) + s_d\varepsilon\mathbf{I}$, where $s_d$ is a dimensional scaling factor; and $\varepsilon$ is a very small value to ensure ergodicity. This adaptive approach enables the proposal function to "learn" from history, gradually approximating the posterior covariance during the sampling process.

However, extending this to a trans-dimensional case presents a challenge, as illustrated in Fig. 3. Specifically, frequent inter-model transitions result in changing dimensions and interpretations of parameters during the sampling process. As exemplified in this figure, "fresh" parameters emerge during inter-model transitions, yet almost no sampling history is available for their adaptation. Consequently, defining an adaptive scheme meaningful for every model becomes essential.

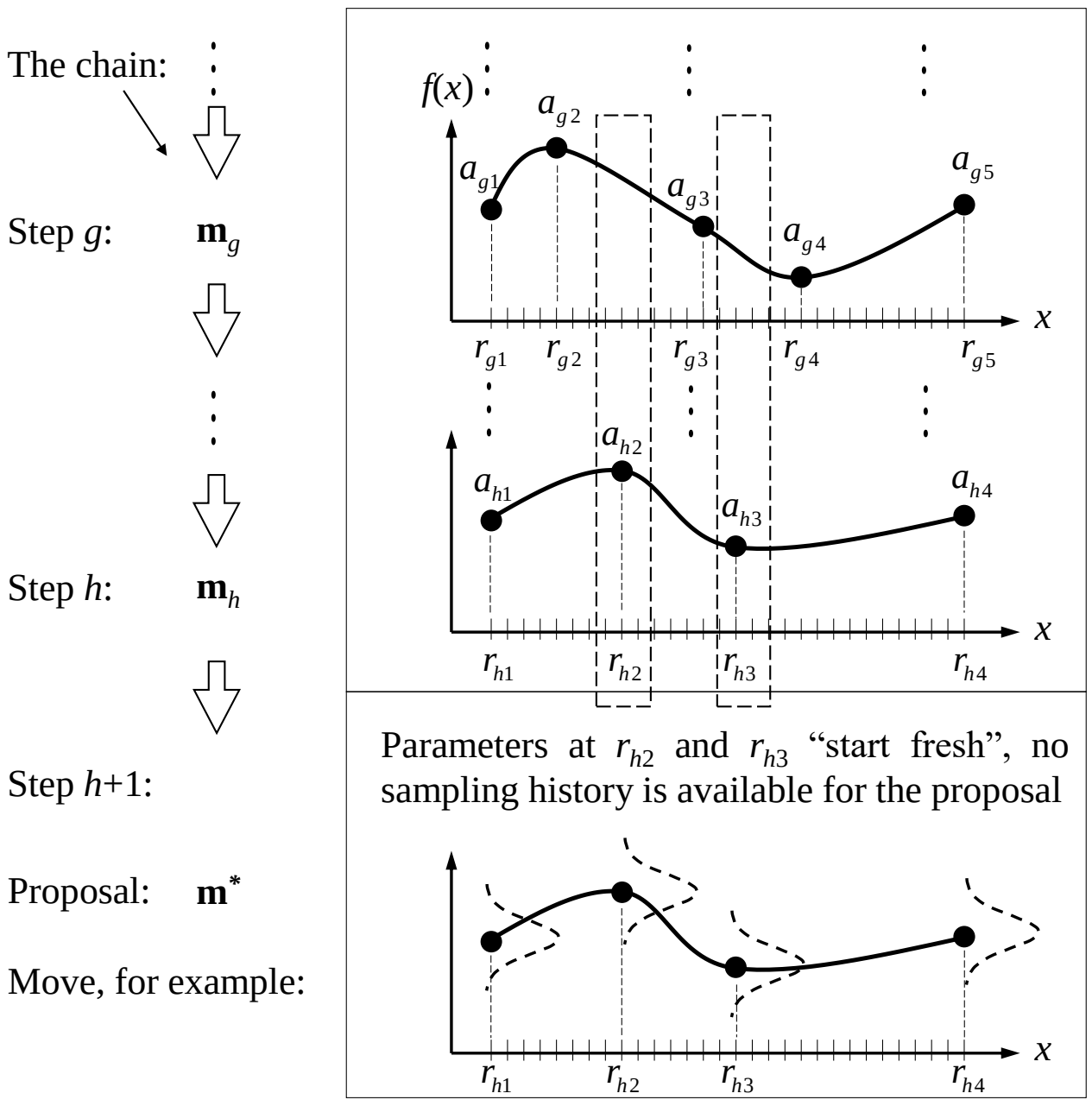

**Fig. 3**. A key issue in extending the adaptive Metropolis algorithm to a trans-dimensional setting: almost no sampling history of the "fresh" parameters is available for adaptation.

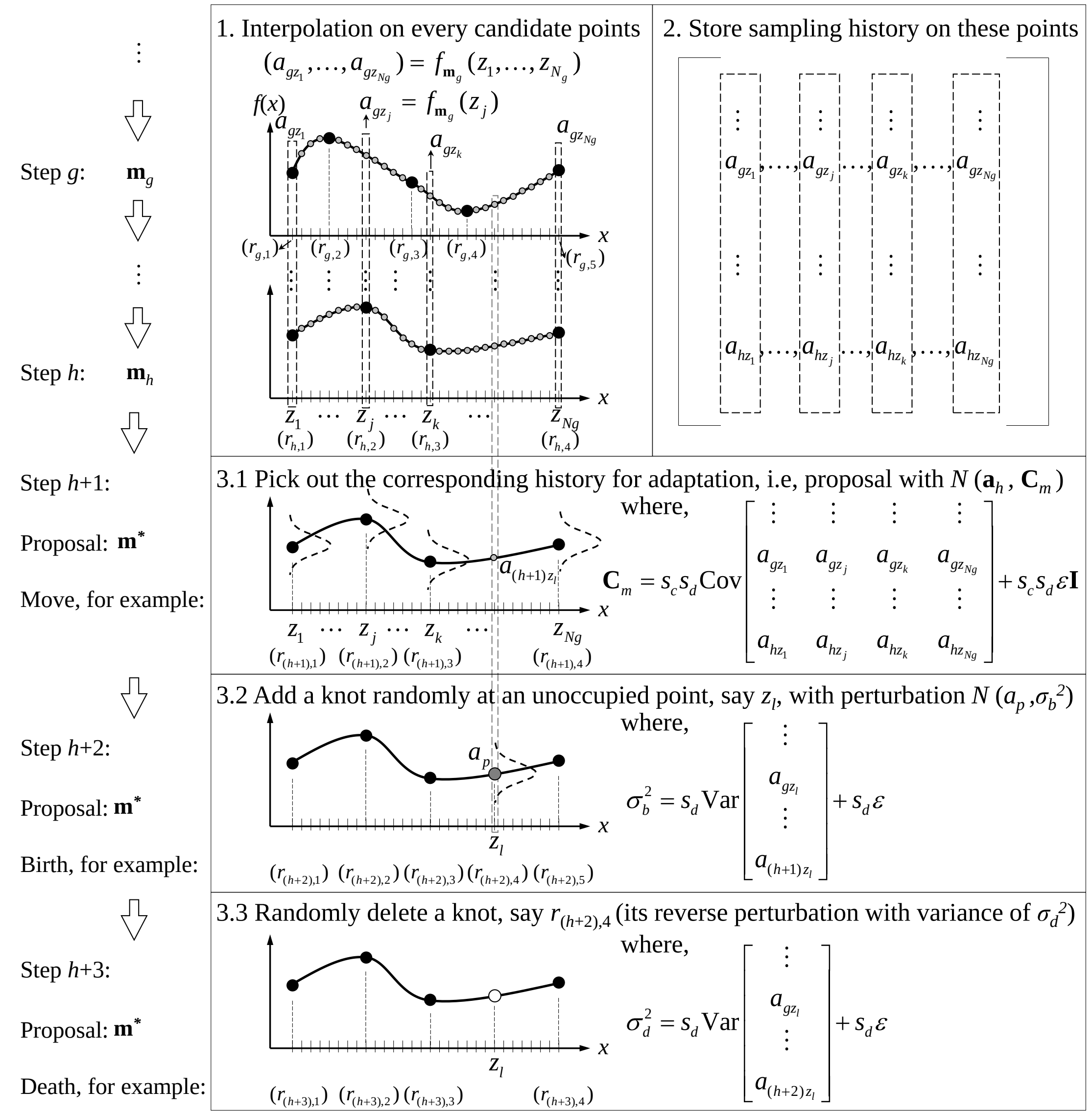

**Fig. 4**. The key ideas of the adaptive proposal RJMCMC. (Note: $s_c$ and $s_d$ represent the dynamic and dimensional scaling factors, respectively, as explained later)

Addressing this challenge, an idea is presented and visually represented in Fig. 4:

i) In each sampling step, compute interpolated values on every candidate point using the function $f_{\mathbf{m}}$ of the current model $\mathbf{m}$. For instance, at step $g$, compute the interpolated value $a_{gzj}$ on the point $z_j$;

ii) Collect and store all these interpolated values at each sampling step, forming the sampling history of the candidate points. For example, on point $z_j$, store the values $(a_{1zj},\ldots,a_{gzj},\ldots,a_{hzj})^{\mathrm{T}}$ from step 1 to the current step $h$.

iii) Activate the sampling history associated with the corresponding candidate points when a forthcoming proposal is related to a specific group of parameters. This sampling history records the history of curve features and can be used to inform an efficient proposal. In Fig. 4, for instance, if a model is at candidate points ($z_1$, $z_j$, $z_k$, $z_{Ng}$), utilize the sampling history of the four points to inform a proper move proposal. This is similar to birth and death proposals, which are also exemplified in this figure.

### 3.2 Implementation

While the concept of adaptive proposal RJMCMC is straightforward, a detailed explanation of its practical implementation is necessary. Let $N_g$ denote the number of candidate points; $\mathbf{j}$ the index vector of these points, $\mathbf{j} = (1,\ldots,N_g)$; and $\mathbf{z_j}$ the coordinates on the points $\mathbf{z_j} = (z_1,\ldots,z_{N_g})$. Since candidate points are generally set to be dense, i.e., $N_g$ is large, storing the sampling history for the entire set of candidate points becomes resource-intensive. Following Haario et al. [22], we adopt a recursive technique. Specifically, we collect the entire covariance matrix related to all the candidate points at an initial period as $\mathbf{C}_{t-1}$ ($N_g \times N_g$); and then update this matrix using the recursive formula:

$$\mathbf{C}_t = \mathbf{C}_{t-1} + \frac{1}{t}\{[f_{\mathbf{m}_t}(\mathbf{z_i}) - \bar{f}_{t-1}(\mathbf{z_i})][f_{\mathbf{m}_t}(\mathbf{z_i}) - \bar{f}_{t-1}(\mathbf{z_i})]^{\mathrm{T}} - \mathbf{C}_{t-1}\}, \tag{10}$$

where, when a new sample $\mathbf{m}_t$ is collected, a vector of interpolated values on all the candidate points can be calculated as $f_{\mathbf{m}_t}(\mathbf{z_i})$; $\bar{f}_{t-1}(\mathbf{z_i})$ is also a vector that stores the mean values on the $N_g$ points during previous $t-1$ steps, and is updated via:

$$\bar{f}_t(\mathbf{z_i}) = \bar{f}_{t-1}(\mathbf{z_i}) + \frac{1}{t}[f_{\mathbf{m}_t}(\mathbf{z_i}) - \bar{f}_{t-1}(\mathbf{z_i})]. \tag{11}$$

Consequently, in forthcoming sampling steps involving parameters located at specific candidate points, the corresponding rows and columns of the covariance matrix can be extracted to generate a proposal variance/covariance. The detailed procedure is summarized as follows.

1. During the initial $t_0$ steps ($t_0$ is small, set to 1000 in the following examples), simulate the chain using conventional RJMCMC. Collect the sampling history for each candidate point through interpolation using the function $f$: $\{f_1(\mathbf{z_i}); \ldots; f_{t_0}(\mathbf{z_i})\}$.

2. At the end of $t_0$ step, compute the initial historical covariance matrix $\mathbf{C}_{t_0} = \mathrm{Cov}\{f_1(\mathbf{z_i});\ldots; f_{t_0}(\mathbf{z_i})\}$ as well as the historical mean vector.

3. After $t_0$ steps, proceed the chain using adaptive proposal RJMCMC:

- For a move step:
  1) identify the index vector (denoted by $\mathbf{c}$) corresponding to the candidate points occupied by the current parameters;
  2) extract the corresponding part from the covariance matrix, i.e., $\mathbf{C}_{t-1}(\mathbf{c},\mathbf{c})$;
  3) generate candidate values $\mathbf{a}^*$ by applying a Gaussian perturbation to $\mathbf{a}_{t-1}$, following $\mathbf{a}^* \sim N(\mathbf{a}_{t-1}, \mathbf{C}_m)$, where $\mathbf{C}_{\mathrm{m}} = s_c s_d\, \mathbf{C}_{t-1}(\mathbf{c},\mathbf{c}) + s_c s_d \varepsilon \mathbf{I}$; $s_{\mathrm{c}}$ and $s_d$ represent the dynamic and dimensional scaling factors, respectively, as explained later;
  4) obtain the candidate sample $\mathbf{m}^*$, where $\mathbf{m}^* = (n^*, \mathbf{r}^*, \mathbf{a}^*)$, with $n^* = n_{t-1}$; $\mathbf{r}^* = \mathbf{r}_{t-1}$. accept this proposal, $\mathbf{m}_t = \mathbf{m}^*$, based on the acceptance ratio with Eq. (6); otherwise, reject the proposal: $\mathbf{m}_{\mathrm{t}} = \mathbf{m}_{t-1}$;
  5) finally, update $\mathbf{C}_{t-1}$ to $\mathbf{C}_t$ using Eqs. (10−11).
- For a birth step.
  1) increase the number of parameters by 1: $n^* = n_{t-1}+1$; add a new knot $r_b$ randomly and uniformly at a candidate point that was not occupied; identify the index number (denoted by $c$) of $r_b$;
  2) pick the corresponding part from the covariance matrix, i.e., $\mathbf{C}_{t-1}(c,c)$;
  3) assign a value, $a_b$, to this new knot by applying a Gaussian perturbation to the

previous interpolated state value $a_p$, with $a_b \sim N(a_p, \sigma_b^2)$, where $\sigma_b^2 = s_d \mathbf{C}_{t-1}$ (c,c)$+s_d\varepsilon$;

4) obtain the candidate sample $\mathbf{m}^*$, where $\mathbf{m}^* = (n^*, \mathbf{r}^*, \mathbf{a}^*)$, with $\mathbf{r}^*=(\mathbf{r}_{t-1}, r_b)$; $\mathbf{a}^*=(\mathbf{a}_{t-1}, a_b)$. accept this proposal, $\mathbf{m}_t = \mathbf{m}^*$, based on the acceptance ratio with Eq. (6); otherwise, reject the proposal: $\mathbf{m}_t = \mathbf{m}_{t-1}$;

5) finally, update $\mathbf{C}_{t-1}$ to $\mathbf{C}_t$ using Eqs. (10−11).

- The death step is almost identical to that in conventional RJMCMC but with its reverse step being the aforementioned birth step, incorporating an adaptive variance.

4. Repeat step 3 until sufficient samples have been collected.

Two tips are provided to enhance the stability and efficiency of the algorithm:

1) Set the dimensional scaling factor $s_d$ to $s_d=(2.4)^2/d$, where $d$ is number of parameters in the proposal. This setting has been shown to optimize the mixing properties of the chain when dealing with a Gaussian posterior and Gaussian proposal [22, 25].

2) Introduce a dynamic scaling factor $s_c$ in the move step. The integration of a dynamic scaling factor is a common strategy in adaptive algorithms to facilitate the learning of the target distribution. This factor is particularly advantageous when the initial estimate of the covariance matrix is a poor guess. In this study, we adopted a powerful technique suggested by Andrieu and Thoms [26], which involves the adaptive modification of $s_c$. This adaptation aims to coerce the acceptance rates of the move step towards a preset target value. The recursive formula for $s_c$ is expressed as follows:

$$\log(s_{c,i}) = \max\{\min\{\log(s_{c,i-1}) + \gamma_i\delta_i, \overline{s}_c\}, \underline{s}_c\}. \qquad (12).$$

where $s_{c,i}$ is the dynamic scaling factor at the $i$-th move step; $\overline{s}_c$ (resp. $\underline{s}_c$) is a very large, e.g., $1\times10^{10}$ (resp. very small, $1\times10^{-10}$) nonnegative number ensuring that $s_c$ is bounded; $\delta_i = \alpha_i - \alpha^*$, where $\alpha_i$ is the acceptance ratio at the $i$-th move step, and $\alpha^*$ is the target acceptance ratio, set at 0.234 [27] for all the following examples; $\gamma_{i+1} = 1/(i+1)^\beta$ according to the recommendation by Andrieu and Moulines [28]. Throughout this paper, $\beta$ is fixed at 0.5 to facilitate an effective adjustment of $s_c$.

Roberts and Rosenthal [29], as well as Haario et al. [22], underscore the importance of demonstrating that adaptive MCMC algorithms possess appropriate ergodic properties. In this context, we include a discussion, confirming that the presented adaptive RJMCMC algorithm adheres to the principle of diminishing adaptation. This principle is a standard sufficient condition for ensuring ergodicity of the sampler [29]. For detailed insights on this discussion, please refer to Appendix B.

### *3.3 The auxiliary-tempered version*

Parallel tempering (PT), serving as an auxiliary method, facilitates the chain's exploration of poorly connected, high-probability regions. Tempered densities enhance the acceptance of inter-model transitions, while swaps enable the un-tempered chain to explore diverse models more frequently.

The implementation of PT is nearly independent of the adaptive proposal

RJMCMC algorithm, requiring only a few minor adjustments. $T$ chains run in parallel with different temperatures. Likelihoods undergo tempering, modifying the acceptance ratio for within-chain sampling as follows:

$$\alpha(\mathbf{m}^* | \mathbf{m}_c) = \min\{1, \frac{p(\mathbf{D}|\mathbf{m}^*)^{Tem_i} p(\mathbf{m}^*) q(\mathbf{m}_c | \mathbf{m}^*)}{p(\mathbf{D}|\mathbf{m}_c)^{Tem_i} p(\mathbf{m}_c) q(\mathbf{m}^* | \mathbf{m}_c)} |\mathbf{J}|\}, \tag{13}$$

where, $Tem_i$ is the temperature of the $i$-th chain, ($i$=1,…,$T$). The adaptive proposal RJMCMC algorithm is independently applied in each chain, with each chain advancing with its own covariance matrix. This necessitates $T$ covariance matrices (or a 3-dimensional tensor, $N_g \times N_g \times T$) for storing covariance history.

Regarding between-chain swaps, they operate independently of the adaptive proposal RJMCMC. Here, an advanced non-reversible swap technique [15] is employed, and the temperatures are adjusted during the burn-in period based on the acceptance rate of swaps, aiming for a range of 0.1 to 0.4 [30]. For a detailed implementation, readers are referred to Syed et al. [15] It's noteworthy that significantly higher acceptance rates could be targeted using non-reversible PT. Pseudocodes for the adaptive proposal RJMCMC (hereafter AP-RJMCMC for simplicity) and the auxiliary-tempered version, i.e., adaptive proposal PT-RJMCMC (hereafter AP-PT-RJMCMC for simplicity), are provided in Appendix C.

## 4 Efficiency examination

Two numerical examples are employed to evaluate the efficiency of the proposed algorithms. The first example is a typical regression problem taken from Denison et al. [6] and Fan et al. [8]. The second example is a specially designed and more challenging case aimed at not only assessing the efficiency but also revealing the underlying mechanisms of the algorithms.

The efficiency of different samplers was compared using the same numerical example. Specifically, we compared AP-RJMCMC against RJMCMC, as well as the auxiliary-tempered version, AP-PT-RJMCMC against PT-RJMCMC. All other parameters were kept identical for different samplers, except for the proposal function, which is detailed in Table 1. Specifically, all samplers were initialized with a function $f(x)$=0 with 4 knots, and ran for $2\times10^6$ steps. The primary distinction was in the proposal functions: AP-RJMCMC and AP-PT-RJMCMC adapted their proposal functions from the sampling history, while RJMCMC and PT-RJMCMC employed a non-adaptive format.

**Table 1**. Parameter settings for the control experiments

| Sampler | Initial state | Chain length | Proposal variance for birth/death, $\sigma_b^2 = s_d\mathbf{C}_{t-1} + s_d\varepsilon$ | Proposal covariance for move, $\mathbf{C}_m = s_c s_d \mathbf{C}_{t-1} + s_c s_d \varepsilon \mathbf{I}$ | Setting for PT |
|---|---|---|---|---|---|
| AP- | $f(x)$=0, | $2\times10^6$ | Adapt $\mathbf{C}_{t-1}$ | Adapt $\mathbf{C}_{t-1}$ using | - |
| RJMCMC | e.g., in Numerical example 1: $\mathbf{m}_0=[n_0,r_0,a_0]$; with $n_0=4$; $r_0=[-2,-1.96,-1.92, 2]$; $a_0=[0,0,0,0]$ | steps | using Eqs. (10−11) | Eqs. (10−11) and adjust $s_c$ with Eq. (12) | |
| RJMCMC | | | $\mathbf{C}_{t\text{-}1}=1$ | $\mathbf{C}_{t\text{-}1}=\mathbf{I}$ and $s_c=1$ | |
| AP-PT-RJMCMC | | | Adapt $\mathbf{C}_{t\text{-}1}$ using Eqs. (10−11) | Adapt $\mathbf{C}_{t\text{-}1}$ using Eqs. (10−11) and adjust $s_c$ with Eq. (12) | 10 parallel chains with non-reversible swap [15] |
| PT-RJMCMC | | | $\mathbf{C}_{t\text{-}1}=1$ | $\mathbf{C}_{t\text{-}1}=\mathbf{I}$ and $s_c=1$ | |

Note: **I** is the identity matrix. It should be noted that setting the covariance matrix as **I** and the variance as 1 may not be advantageous for the conventional algorithms. This is primarily attributed to the difficulty in guessing a more appropriate value, an inherent challenge in these algorithms, particularly pronounced in diverse application contexts.

Monitoring convergence is essential for evaluating the efficiency of samplers. Establishing a robust convergence diagnosis for trans-dimensional samplers is typically a challenging task. In this study, we employ a pragmatic convergence diagnosis method [18]. The method involves running two instances of an identical sampler independently and concurrently, collecting samples from each to monitor related statistics. The convergence criteria are considered to be reached when statistics from the two runs agree with sufficient precision. While not foolproof for rigorous diagnosis, a direct comparison between two distinct samplers provides an objective means to indicate the efficiency superiority of one over the other. The selected statistics are the mean and standard deviation of the function, denoted by $R_{c1}$ and $R_{c2}$ respectively:

$$R_{c1}=\frac{1}{N_g}\sum_{i=1}^{N_g}\frac{|\mu_{i1}-\mu_{i2}|}{(\sigma_{i1}+\sigma_{i2})/2}, \tag{14}$$

$$R_{c2}=\frac{1}{N_g}\sum_{i=1}^{N_g}\frac{|\sigma_{i1}-\sigma_{i2}|}{(\sigma_{i1}+\sigma_{i2})/2}, \tag{15}$$

where $\mu_{i1}$ and $\sigma_{i1}$ denote the mean and standard deviation of the function values on candidate point $i$ during a monitoring step in the first run; $\mu_{i2}$ and $\sigma_{i2}$ denote the same statistics in the second run. In the sampling process, convergence criteria are considered met only when both $R_{c1}$ and $R_{c2}$ are less than 0.2 [18]. Further details on the indicators can be found in the referenced literature. It is important to note that the sampler can exhibit variability from run to run due to the inherent randomness of the sampling process. To address this variability and ensure a comprehensive convergence analysis, each sampler underwent 20 independent runs. Subsequently, each run was systematically paired with every other run (excluding self-pairing), resulting in a total of 190 pairs of 'two-run' instances for convergence diagnosis.

## 4.1 *Numerical example* 1

### 4.1.1 *Preliminaries*

A dataset is created by introducing noise to a pre-defined function. The objective is to find a regression function (piecewise spline) to fit these data. The true function is $f(x) = \sin(2x) + 2\exp(-16x^2)$ ($x \in [-2, 2]$), and the noise is generated from a zero-mean Gaussian distribution with a standard deviation of 0.3. Fig. 5 illustrates the true function and the corresponding data. As discussed in Introduction, for this regression problem, RJMCMC can be designed using conjugate priors for the coefficients, resulting in significantly improved sampling efficiency [8]. However, we deliberately choose not to employ conjugate priors to showcase the generality of our algorithms. Specifically, this regression problem is treated as a special inverse problem with $g(f, x_i)=f(x_i)$, where all the parameters are jointly sampled. This decision is motivated by the impracticality of using conjugate priors for many real-world inverse problems, especially those lacking closed-form forward models for analytically integrating coefficients out of the posterior.

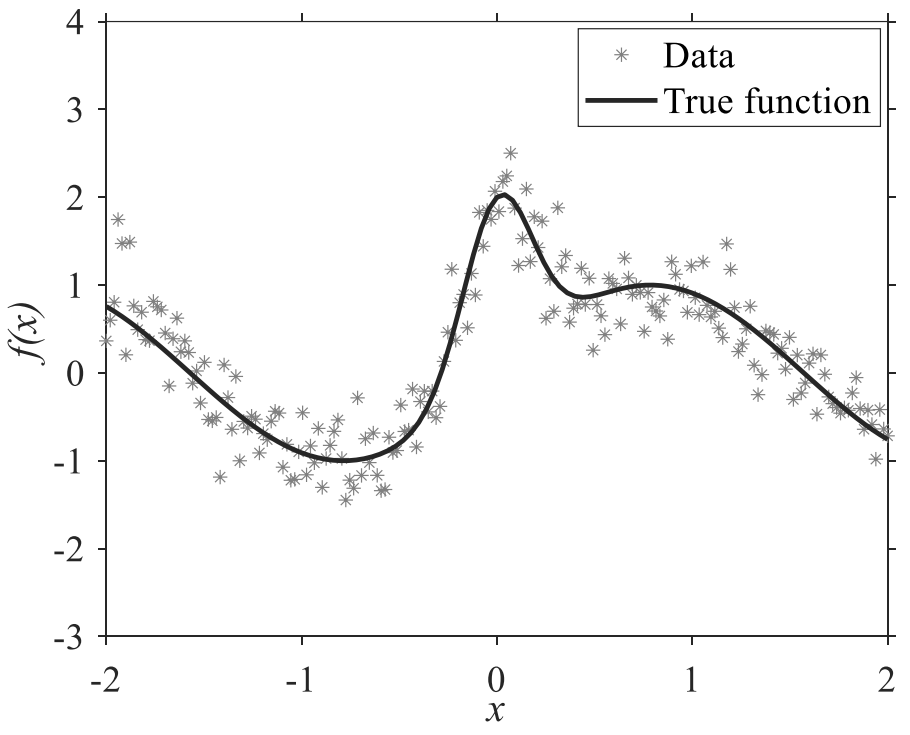

**Fig. 5**. Data for curve fitting used in numerical example 1.

We evenly place 101 candidate points within the data domain, i.e., $N_g = 101$, as discussed during the analysis of the results. Uniform priors were applied to the parameters $\mathbf{m} = (n, \mathbf{r}, \mathbf{a})$, i.e., $n \sim$ Uniform (2, 101), where 2 is a minimum number of knots to fit a curve and 101 is the maximum number of knots when all candidate points are occupied; the conditional prior of $p(\mathbf{r}|n)$ is determined as the distribution of $n$–2 points sampled without replacement from the set of interior grid points, with a probability mass of $1/C(N_g\text{–}2,n\text{–}2)$; the prior for $\mathbf{a}$ is taken to be independent and $a_i \sim$ Uniform (–10, 10) ($i = 1,\dots,n$), where the bounds [–10, 10] cover a broader range than the dataset, which is around [–2,3]. As discussed in section 2.2, the noise level is assumed to be known in advance and $\bar{\sigma}_e$ in the likelihood function is taken to be 0.3.

### 4.1.2 *Results*

The representative sampling process and inversion results of the four samplers are presented in Fig. 6. The first two columns of this figure display the sampling history of data fit (log-likelihood, *LL*) and the number of knots ($n$), respectively. The zoomed-in boxes in the first column highlight the last 25% of the samples' *LL*. As evident in the boxes, RJMCMC exhibited relatively poor performance, characterized by its slow movement around the model space, whereas AP-RJMCMC demonstrated superior mixing capabilities. Furthermore, parallel tempering showed improved mixing compared to RJMCMC, while AP-PT-RJMCMC moved through the model space much more rapidly than PT-RJMCMC.

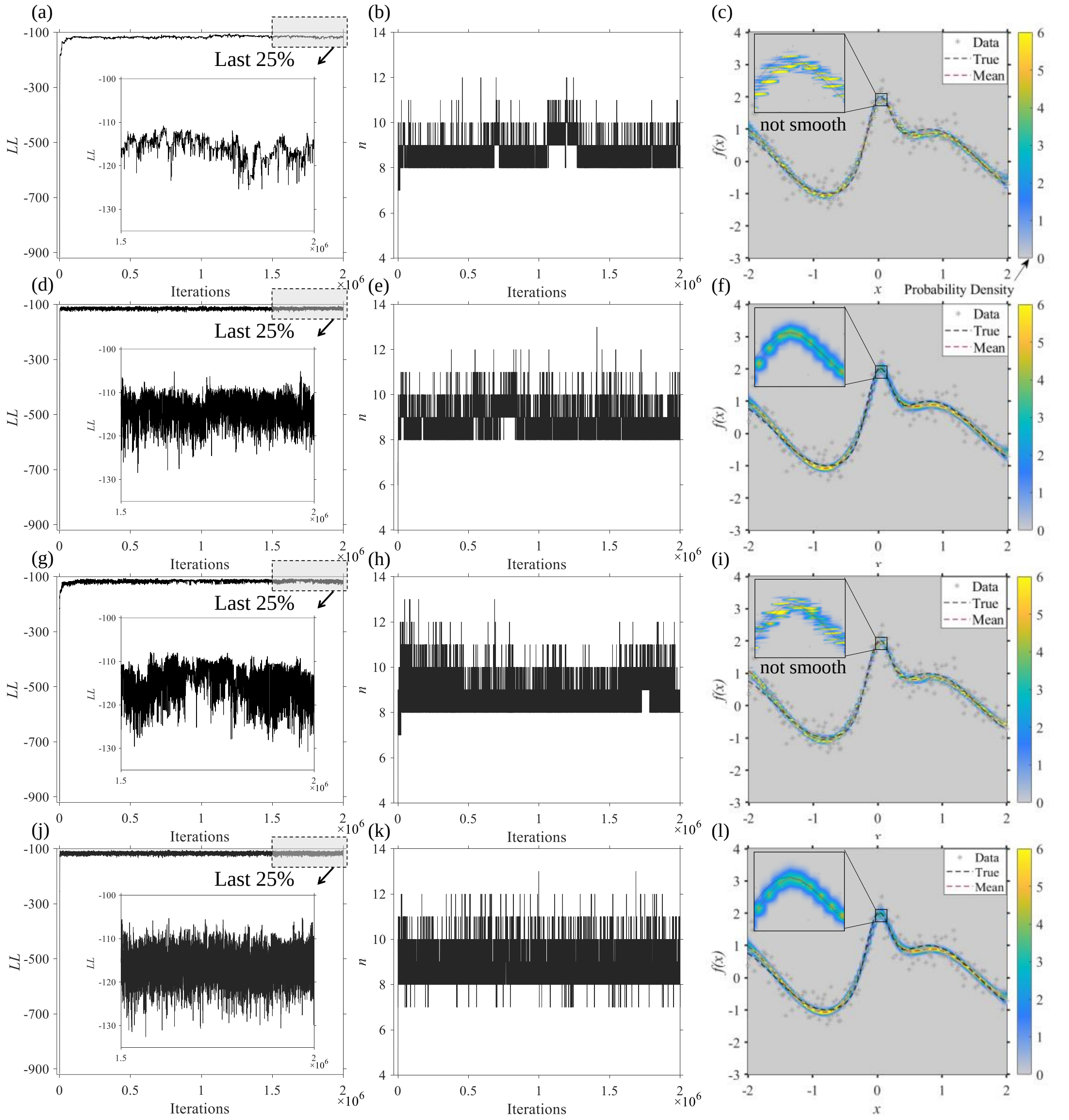

**Fig. 6**. Sampling history and inversion results in numerical example 1: (a)–(c) RJMCMC; (d)–(f) AP-RJMCMC; (g)–(i) PT-RJMCMC; (j)–(l) AP-PT-RJMCMC. The coloms, from left to right, show the sampling history of data fit (*LL*), number of knots (*n*), and inversion results, respectively.

The posterior distribution of *f* was estimated using the second half of the chain. The marginal probability densities, along with the posterior means of the function, were plotted to the right of the chains. As depicted in the results from AP-RJMCMC and AP-PT-RJMCMC, the true function is closely encompassed by high-probability regions. In addition, the posterior means perform a smooth fit to the true function, suggesting effectiveness of these algorithms. This also indicates that the chosen grid density ($N_g$ being 101) is sufficient for the numerical example. While this criterion for evaluating the appropriateness of $N_g$ will be similarly applied in subsequent examples, further discussion on this aspect will not be presented hereafter. The posterior means of RJMCMC and PT-RJMCMC also show a reasonable alignment with the true function and data; however, their posterior distributions are less smooth, particularly around $x$=0, where the curve exhibits a sharp peak, indicating potential poor mixing.

Most notably, Fig. 7 summarizes the convergence lengths for the four samplers. These lengths, evaluated using the predetermined criteria involving indicators $R_{c1}$ and $R_{c2}$, are based on the entire set of 190 pairs. Overall, AP-RJMCMC demonstrated superior efficiency and stability compared to RJMCMC. Specifically, only 3 out of 190 pairs processed by RJMCMC met the convergence criteria within $200\times10^4$ steps, with an average length of $179.4\times10^4$ steps. In contrast, all 190 pairs processed by AP-RJMCMC satisfied the convergence criteria, with a significantly reduced average convergence length of $37.4\times10^4$ steps. Furthermore, it was observed that parallel tempering facilitated sampling convergence: 34 out of 190 pairs processed by PT-RJMCMC successfully met the criteria, with an average length of $178.3\times10^4$ steps. However AP-PT-RJMCMC demonstrated the most remarkable performance, achieving the convergence criteria in an average of just $8.2\times10^4$ steps.

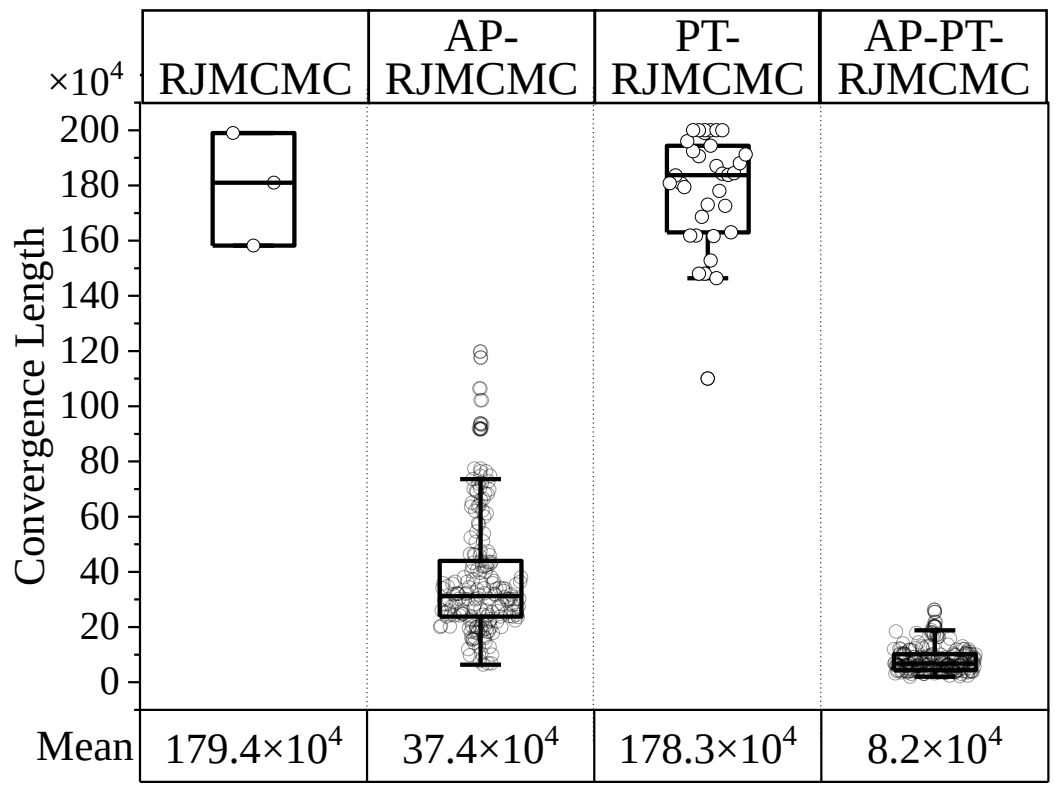

**Fig. 7**. Summary of the convergence lengths of the four samplers (determined using the specified criteria involving indicators $R_{c1}$ and $R_{c2}$) based on the 190 pairs in numerical example 1.

It is crucial to recognize that the implementation of adaptive algorithms and parallel may incur increased computational demands. For a comprehensive comparison of efficiency, we recorded the execution time of the samplers. These simulations were executed in MATLAB on a desktop with a Ryzen 9, 12-Core, 3.8GHz Processor. In terms of the mean convergence lengths presented in Fig. 7, AP-RJMCMC required 154.2 s, while RJMCMC took 577.9 s. Similarly, AP-PT-RJMCMC took 306.2 s, a substantial reduction from the 5459.7 s necessitated by PT-RJMCMC. These results suggest that the adaptive process does not notably increase computational costs; rather, adaptive algorithms manifest as markedly more efficient than the conventional ones. However, it is noteworthy that AP-PT-RJMCMC required 98.6% additional time to meet the convergence criteria compared to AP-RJMCMC. Although this increase in time consumption might initially appear disadvantageous, it is crucial to highlight that the parallel tempering framework was not executed within a parallel computing environment in this paper. The efficiency of this framework could be enhanced through parallel execution approaches, such as GPU-based techniques [31]. Furthermore, subsequent sections of this paper will present specific scenarios where the integration of parallel tempering is essential.

We also employed the Autocorrelation Function (ACF) to evaluate the efficiency of the four samplers. Specifically, we selected representative instances from the estimated convergence pairs depicted in Fig. 7. The sampling trace at each candidate point was recorded using the interpolation technique introduced in Fig. 4. These interpolated sampling traces were then utilized to plot the ACF. Fig. 8 displays the ACF at the first candidate point for the four samplers (the ACF trends at the other candidate points were observed to be similar to those in Fig. 8). Notably, the ACF for the adaptive algorithms demonstrated a more rapid decline compared to the conventional ones. This observation reinforces the indication of enhanced mixing and a higher convergence rate for the adaptive algorithms.

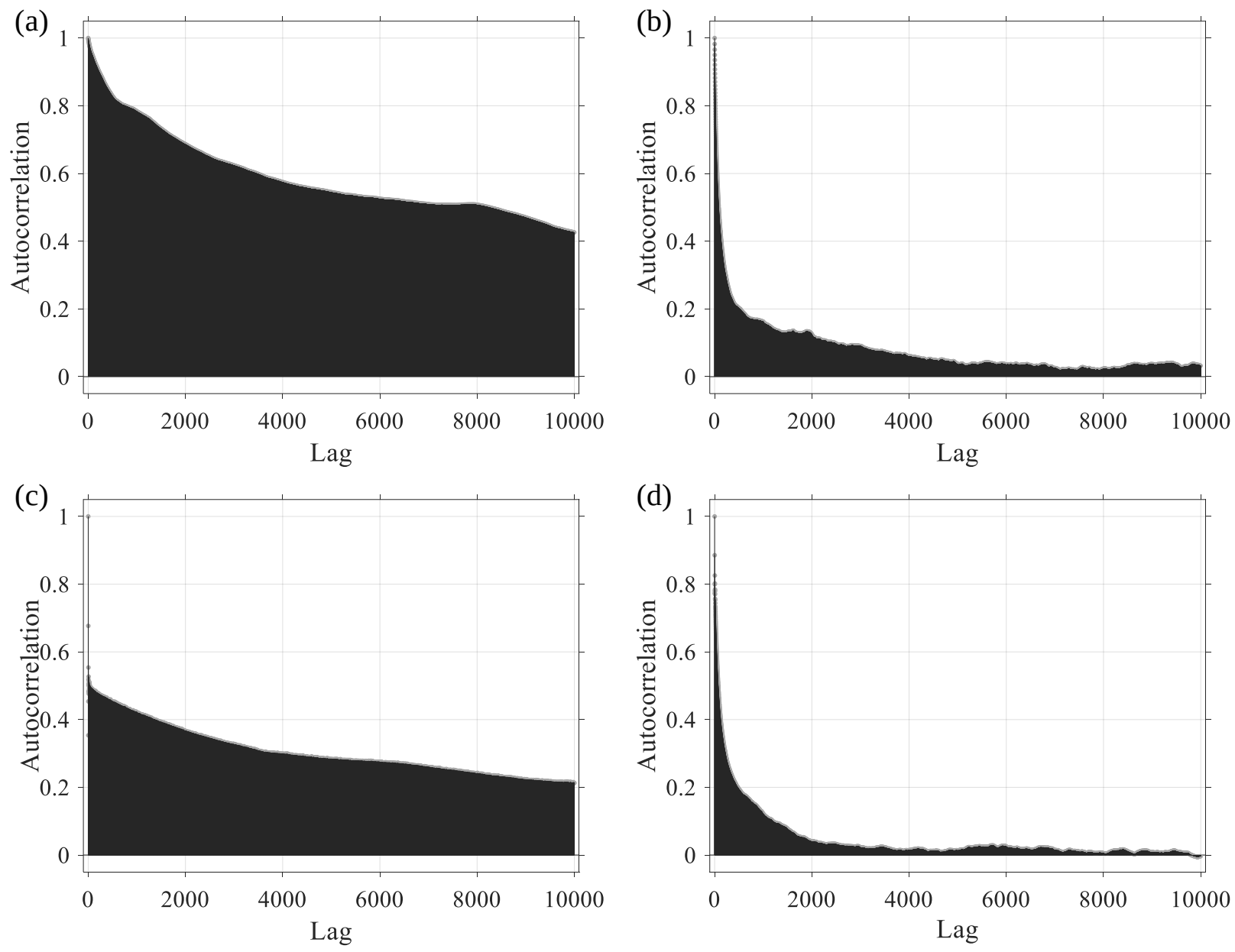

**Fig. 8**. Autocorrelation function of the four samplers at the first candidate point: (a) RJMCMC; (b) AP-RJMCMC; (c) PT-RJMCMC; (d) AP-PT-RJMCMC.

## 4.2 *Numerical example* 2

### 4.2.1 *Preliminaries*

The objective is to recover a piecewise constant function based on the data presented in Fig. 9. Although this problem may seem innocuous, the posterior distribution proves challenging due to the limited data. Specifically, as exemplified in this figure, both potential lines 1 and 2 align equally well with the data. This alignment is attributed to the fact that values at a knot of a function in the transition zone remain unaffected by the data. Consequently, the uncertainty in inferring the function within the transition zone can be higher than within the pieces (a detailed explanation will be provided in section 4.2.3). This suggests that the posterior distribution of the function can vary significantly along the definition domain, necessitating a more flexible proposal for efficient sampling.

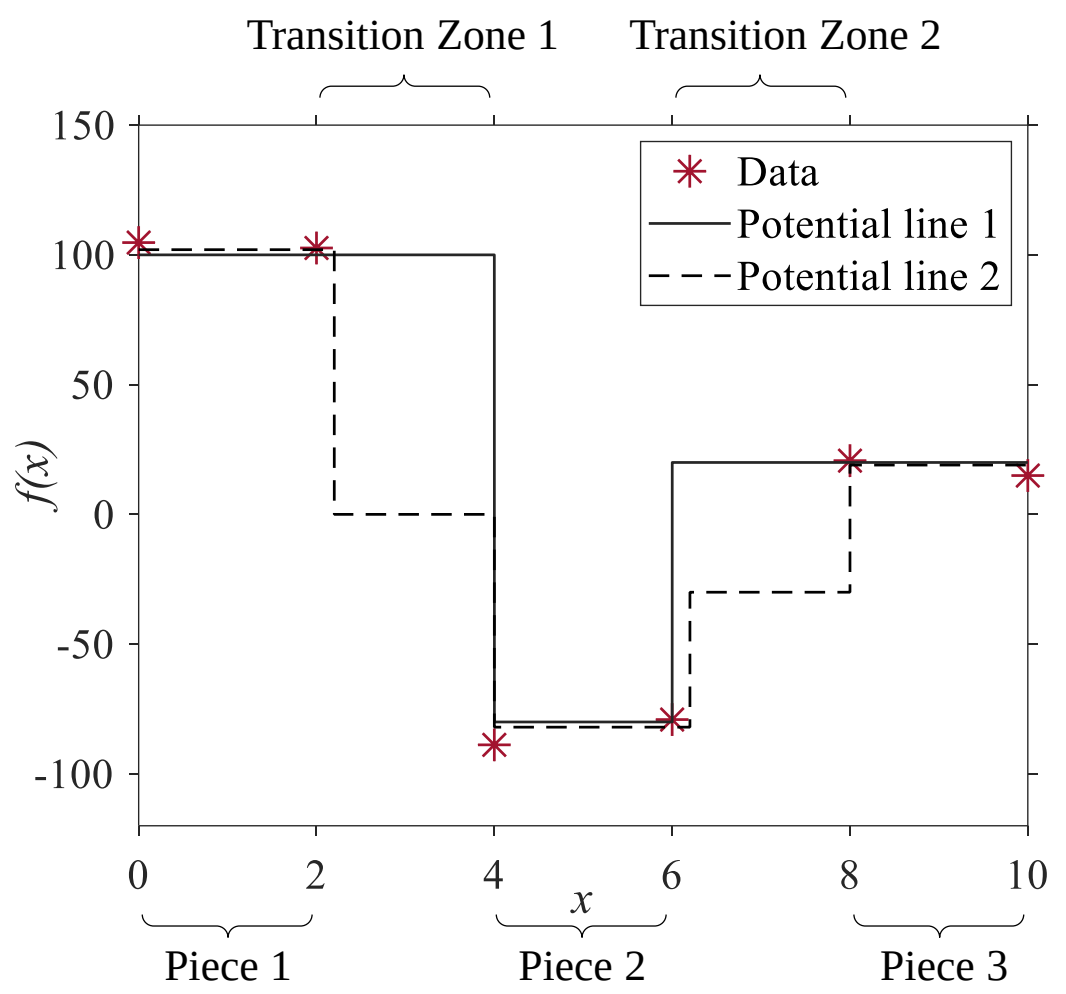

**Fig. 9**. Data for curve fitting used in numerical example 2.

Similarly, 101 candidate points were evenly placed within the definition domain of the data ($N_g = 101$). Uniform priors were applied to the parameters $\mathbf{m} = (n, \mathbf{r}, \mathbf{a})$, i.e., $n \sim$ Uniform (2, 101); $p(\mathbf{r}\,|n)$ is established with a probability mass of $1/C(N_g–2,n–2)$; the prior for $\mathbf{a}$ is considered independent and $a_i \sim$ Uniform (–300, 300) ($i = 1,\ldots,n$), encompassing a broader range than the dataset, which is approximately [–100, 100]. The noise level is assumed to be known in advance, and $\bar{\sigma}_e$ in the likelihood function is set to be 5 according to the data setting. For simplicity, only the advanced version, i.e., the auxiliary-tempered version was tested here, i.e., AP-PT-RJMCMC vs PT-RJMCMC.

### *4.2.2 Results and discussion*

Fig. 10(a) illustrates the monitoring process of $R_{c1}$ and $R_{c2}$ for a representative pair run by PT-RJMCMC. Observationally, $R_{c1}$ and $R_{c2}$ failed to converge to the threshold throughout the entire sampling process, remaining above 0.2 in the later stages. Similar to this representative pair, none of the other 189 pairs run by PT-RJMCMC met the specified convergence criteria within $2\times10^6$ steps.

With the second half of the samples generated via PT-RJMCMC, we plotted marginal posterior densities as well as the posterior means of the function. As shown in Fig. 10(b), although the data are tightly confined within the "hot area", noticeable local fluctuations and instability are observed in the posterior means. To explore this observation further, we rescaled the color density range to highlight relatively lower probability areas in Fig. 10(c). Notably, local modes are evident in this figure, serving as a strong indicator of non-convergence. For example, when a new knot is created in the transition zone, the value on this knot is expected to move freely within the priors, without constraints from the data. Consequently, the posterior density should exhibit an even distribution in this area (refer to section 4.2.3 for a detailed explanation). However, the existence of local modes suggests that the sampler was locally trapped, impeding effective mixing across all low probability areas. This phenomenon ultimately leads to a lack of convergence and contributes to instability in the posterior means.

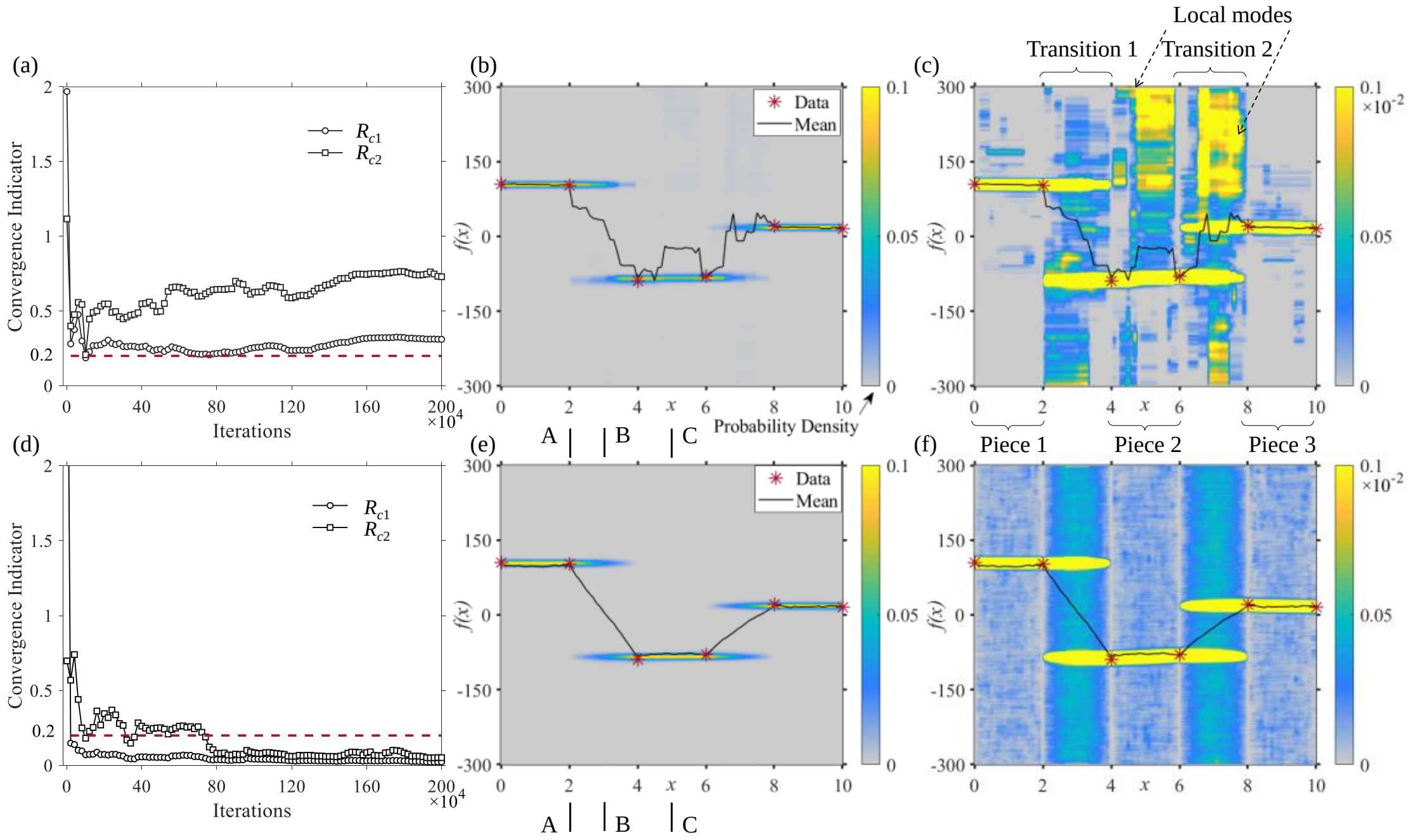

**Fig. 10.** Convergence monitoring and inversion results of the samplers in numerical example 2: (a) convergence monitoring of a representative pair run by PT-RJMCMC; (b)–(c) inversion results by PT-RJMCMC; (d) convergence monitoring of a representative pair run by AP-PT-RJMCMC; (e)–(f) inversion results by AP-PT-RJMCMC.

In contrast, the convergence monitoring of a representative pair run by AP-PT-RJMCMC is presented in Fig. 10(d). Observationally, the sampler met the requirement of convergence around the $75\times10^4$ step, and the indicators remained below the threshold thereafter. Similarly, the inversion results and the rescaled version are shown in Fig. 10(e)–(f). It is evident that the posterior means are considerably smoother and align well with the data. Moreover, no local modes exist, and the densities are evenly distributed in areas where the inference is uninformative from the data. That is, the sampler mixed well in the low probability areas, avoiding getting trapped as seen in PT-RJMCMC.

The effectiveness of AP-PT-RJMCMC in sampling can be attributed to its adaptive proposal function. Specifically, the posterior distribution in this example exhibits significant variations across the definition domain. For the conventional algorithm, identifying an appropriate proposal function that approximates the *a priori* unknown posterior distribution is challenging. However, in AP-PT-RJMCMC, the entire sampling history is recorded, enabling the gradual approximation of the proposal function to the complex posterior covariance. This attribute renders AP-PT-RJMCMC an efficient sampler. To illustrate this further, the recorded variance at sections A, B, and C (shown in Fig. 10e as an example) along the chain is presented in Fig. 11. It is evident that, along with the sampling process, the recorded variance at these sections adapted to approximate their respective posterior variance automatically and gradually. This implies that when a new knot is proposed on a specific section, the proposal on this knot can be informatively adjusted to fit the corresponding posterior variance on that section. This adaptability is considered the most appealing feature of the adaptive algorithm, distinguishing it from the conventional one.

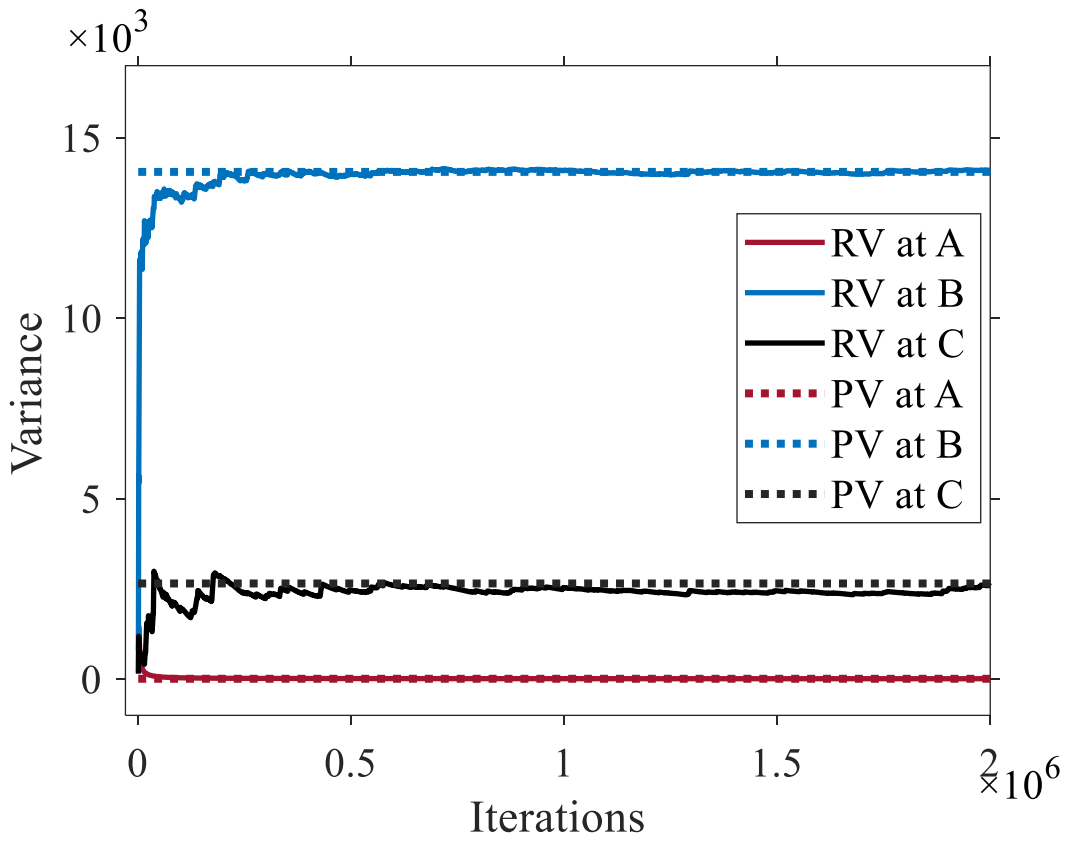

**Fig. 11.** Recorded sampling history variance and the posterir variance at sections A, B, and C by AP-PT-RJMCMC. (RV=Recorded Variance; PV=Posterior Variance)

In addition to the representative pairs analyzed above, a summary of the convergence lengths (determined using the specified criteria involving indicators $R_{c1}$ and $R_{c2}$) for all 190 pairs run by AP-PT-RJMCMC is presented in Fig. 12. It is noteworthy again that PT-RJMCMC failed to meet the criteria within $2\times10^6$ steps across the entire set of 190 pairs. In contrast, all the 190 pairs subjected to AP-PT-RJMCMC successfully met the convergence requirement within $2\times10^6$ steps, with an average length of $53.2\times10^4$. Overall, it is concluded that AP-PT-RJMCMC performed well in achieving a stable posterior distribution in this challenging test while PT-RJMCMC failed.

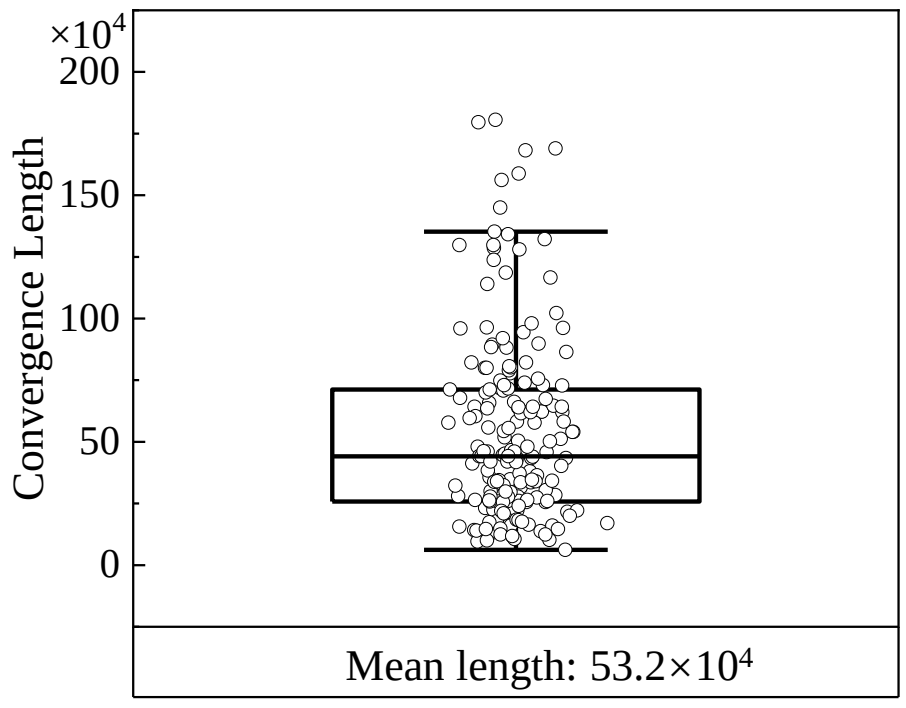

**Fig. 12**. Summary of the convergence lengths of AP-PT-RJMCMC (determined using the specified criteria involving indicators $R_{c1}$ and $R_{c2}$) for the 190 pairs in numerical example 2 (Note: PT-RJMCMC failed to meet the convergence criteria in any of the 190 pairs, resulting in the absence of data for displaying its convergence length).

#### *4.2.3 Discussion on the contribution of parallel tempering*

The comparison between AP-PT-RJMCMC and PT-RJMCMC has highlighted the advantage of adaptive proposal (AP). However, the contribution from PT in AP-PT-RJMCMC cannot be disregarded, as will be demonstrated in the subsequent analysis. AP-RJMCMC was run for numerical example 2 with the same parameters given in Table 1. The results (Fig. 13a) show that the posterior means are smooth and align well with the observed data. However, the posterior distribution (presented in Fig. 13b) appears to be somewhat different from that in Fig. 10(f). Specifically, the presence of local modes indicates a lack of convergence in this instance by AP-RJMCMC.

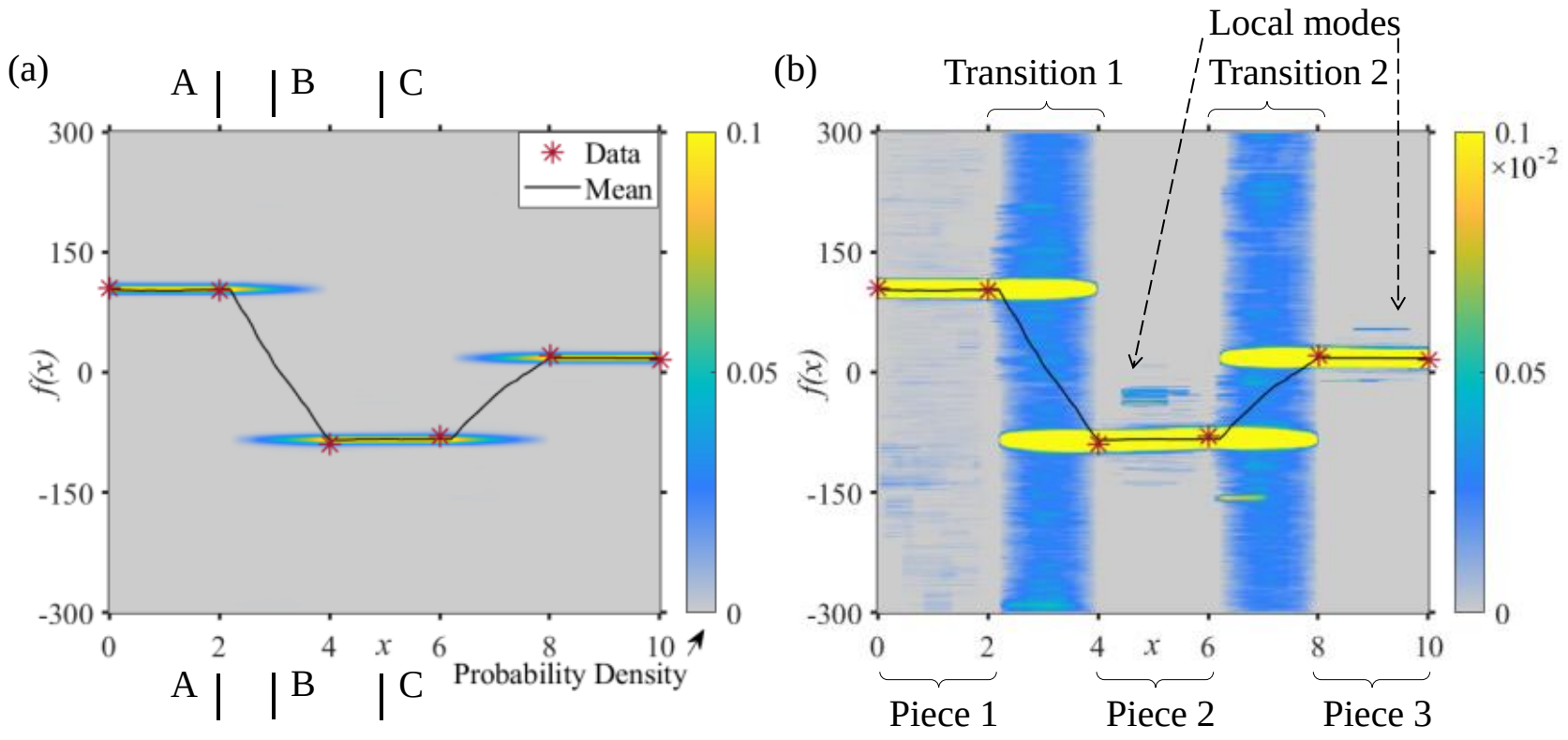

**Fig. 13.** Inversion results by AP-RJMCMC in numerical example 2.

To provide a clearer explanation, a sketch of how the sampler jumps between different models is presented. As shown in Fig. 14(a), if a new knot is proposed in the transition zone ("birthing once"), due to no constraints from the observed data, values on this knot can move freely within the prior since the likelihood is not affected. However, as shown in Fig. 14(b), only when two knots are created within a piece zone ("birthing twice"), can the values on one of the knots move freely. It should be noted that acceptance ratio for a birth step is not only determined by the quality of a proposal but also penalized by the "natural parsimony" (the prior) of RJMCMC. Thus, despite the proposal being adapted, the probability to accept a birth step may still be low, and so the probability of having two nodes in a single piece ("birthing twice") may be significantly lower. It explains why the posterior distribution varies significantly along the definition domain (Fig. 10f). Most importantly, "birthing twice" becoming a rare event causes AP-RJMCMC to fail in achieving ergodic sampling over all the piece zones. By contrast, with the auxiliary chains, PT helps mixing between different models that makes up for the inter-model transition difficulties encountered by AP-RJMCMC. Consequently, AP-PT-RJMCMC mixed well in all the low probability piece zones (Fig. 10f). It is an appealing complement to AP-RJMCMC.

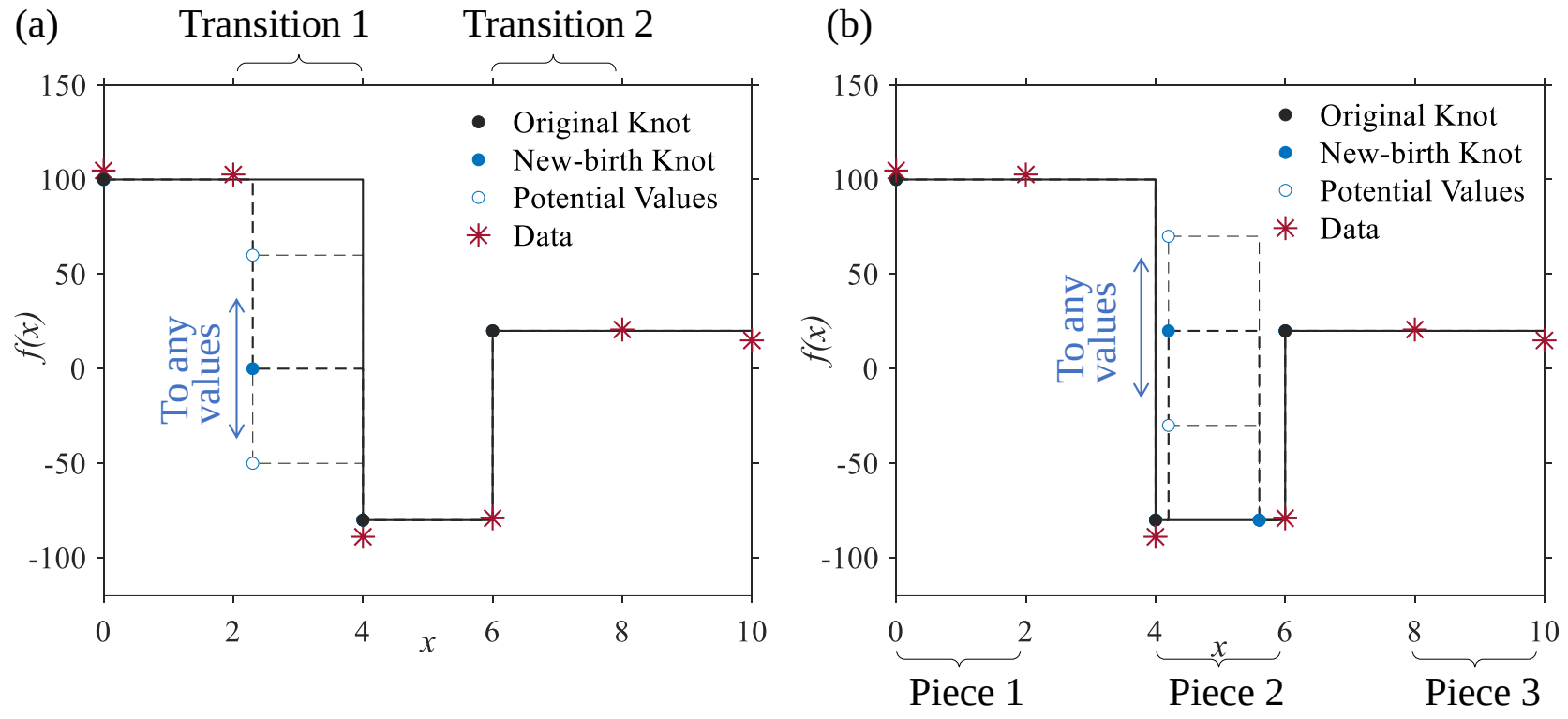

**Fig. 14.** Sketch on how the sampler jumps between different models: (a) creating a knot in the transition zone; (b) creating two knots in the piece zone.

In addition, the wide exploration capacity of hot chains and swap techniques helps the target chain access the high probability areas early and often, which allows the chain to approximate the posterior variance more accurately and quickly. This can be seen by comparing Fig. 11 and Fig. 15. In AP-RJMCMC (Fig. 15), the "learning" of variance at section B appears to be smoother and slower than that in AP-PT-RJMCMC (Fig. 11). In addition, the absence of mixing in the low-probability areas led to an incorrect learning of the variance in section C. Overall, these factors underscore the importance of introducing the auxiliary-tempered version of the adaptive proposal algorithm.

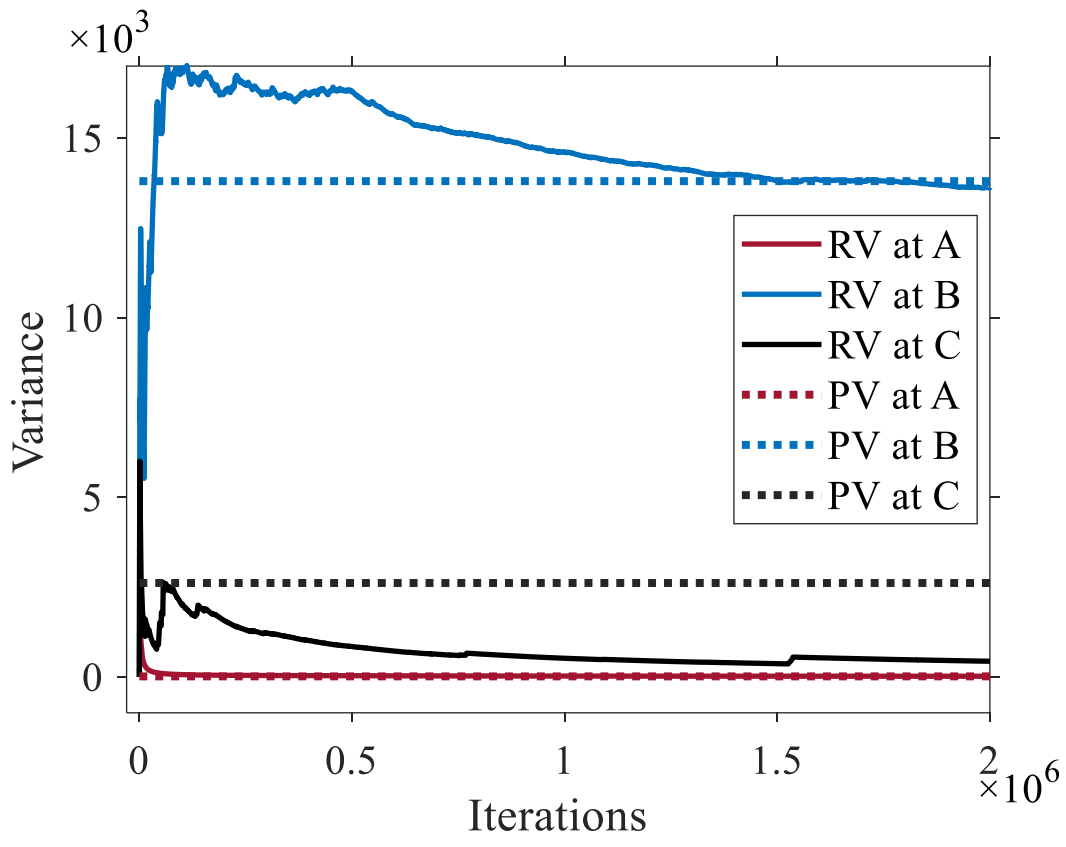

**Fig. 15** Recorded sampling history variance by AP-RJMCMC and the posterior variance by AP-PT-RJMCMC at sections A, B, and C. (RV=Recorded Variance; PV=Posterior Variance)

## 5 Application in engineering practice

### *5.1 Preliminaries*

Similar to numerical example 2, the posterior distribution in practical inverse problems is generally complex and requires an efficient sampler with a proposal function that can "learn" to approximate the *a priori* unknown posterior distribution. Unlike the previous examples ($g(f, x_i)=f(x_i)$), the forward model in a specific inverse problem is determined case by case. Here, we apply the samplers to a field case in geotechnical engineering [13]. As shown in Fig. 16(a), a pile is bent down the slope due to a landslip. The deformations on the pile were recorded by Smethurst and Powrie [32], as illustrated in Fig. 16(b). Identifying the pressures on an underground structure is crucial for its safety assessment. Consequently, the objective is to invert the lateral net pressures on the pile (Fig. 16c) based on the deformation data (Fig. 16b).

A piecewise linear function is employed to fit the actual pressures on the pile. The elastic beam model is adopted as the forward model, calculating the deformation on a concrete pile under any given pressures. A detailed introduction to the forward model can be found in Appendix D. Geometric and mechanical parameters for the forward model were recorded by Smethurst and Powrie [32] and are presented in Fig. 16(a).

101 candidate points were uniformly placed within structural domain of the pile. The inversion of pressures on engineering structures can be ill-conditioned. For such cases, a Poisson prior is typically applied to the number of knots [6–7], and consequently we set $n \sim$ Poisson (22). Uniform priors are employed for $\mathbf{r}$ and $\mathbf{a}$, i.e., the conditional prior of $p(\mathbf{r}\,|n)$ is established with a probability mass of $1/C(N_g–2,n–2)$. For redundancy, we set the prior bounds for $\mathbf{a}$ as the maximum self-weight at the pile bottom: 19 kN/m$^3$ × 14.5 m ≈ 300 kPa [33]. A detailed discussion on this choice can be found in the reference [13]. Accordingly, priors for $\mathbf{a}$ are considered independent and $a_i \sim$ Uniform (–300, 300) ($i$ = 1,…,$n$). Considering the precision of the measurement instrument for deformations, $\bar{\sigma}_e$ in the likelihood function is taken to be 1 mm [32].

Again, the primary focus here is to examine the efficiency of the adaptive proposal algorithm, and only the advanced version was tested here, i.e., AP-PT-RJMCMC vs PT-RJMCMC. For controlled experiments, all the parameter settings are the same as those in examples 1 and 2 given in Table 1. Similarly, each sampler underwent 20 independent runs, generating190 pairs for convergence diagnosis.

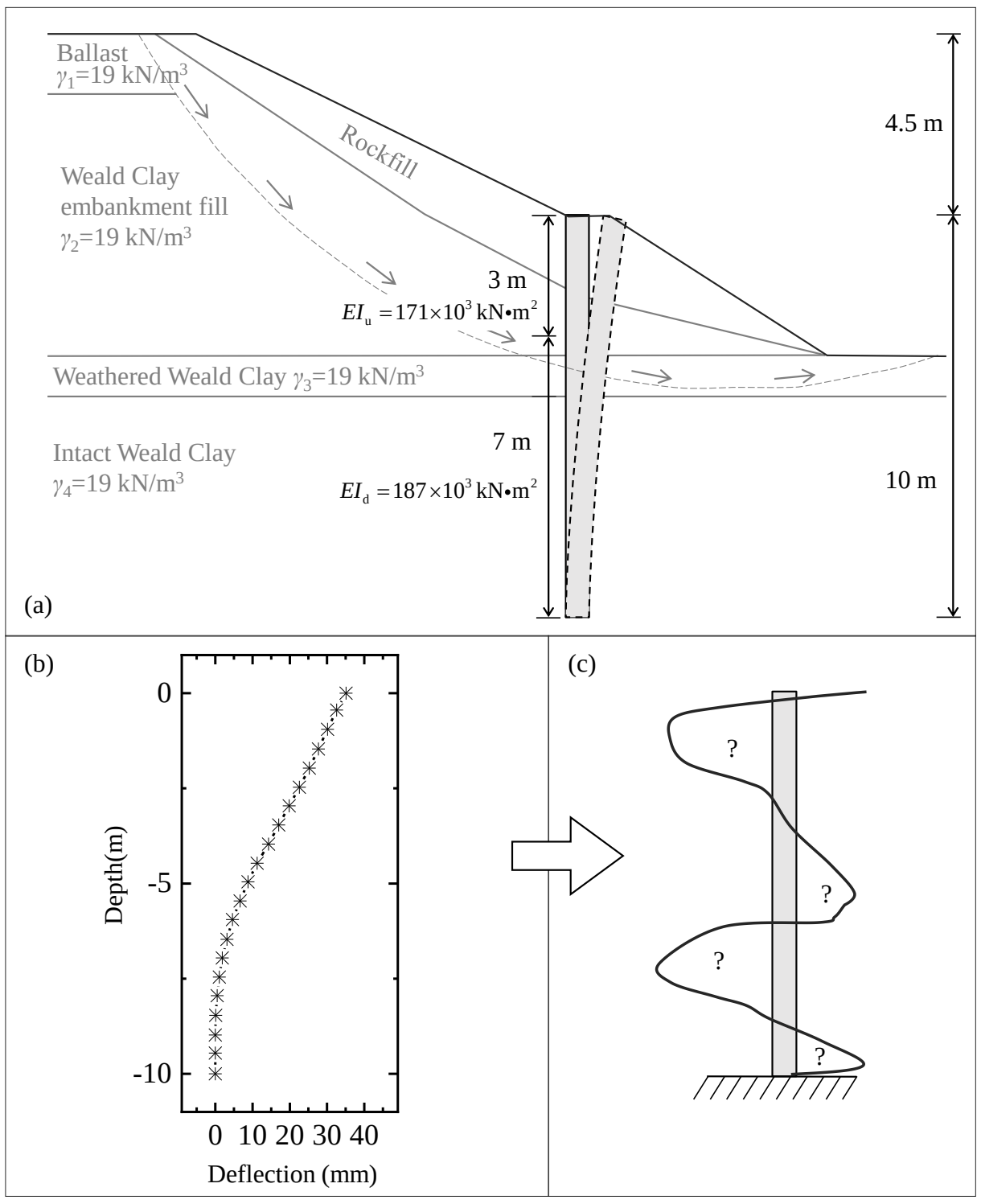

**Fig. 16.** The field example recorded by Smethurst and Powrie [13, 32]: (a) a pile bends down a slope; (b) observed deformations on the pile; (c) inversion of the unknown pressures.

## *5.2 Results*

The monitoring process of $R_{c1}$ and $R_{c2}$ of a representative pair, run by PT-RJMCMC, is presented in Fig. 17(a). Once again, $R_{c1}$ and $R_{c2}$ failed to converge to the threshold throughout the entire sampling process. Similar to this representative pair, none of the other 189 pairs run by PT-RJMCMC reached the convergence criteria within $2\times10^6$ steps. With the second half of the samples, marginal posterior densities and the posterior means of the function were plotted. As shown in Fig. 17(b), neither the posterior distribution nor the posterior means aligned closely with the actual recorded pressures deduced from the strain gauge data [13, 32]. In contrast, all 190 pairs run by AP-PT-RJMCMC successfully reached the specified criteria, similar to the representative pair monitored in Fig. 17(c). Additionally, as illustrated in the inversion results (Fig. 17d), the actual recorded pressures are confined by the “hot areas” and align well with the posterior means.

Furthermore, a summary of the convergence lengths (determined using the specified criteria involving indicators $R_{c1}$ and $R_{c2}$) for all 190 pairs run by AP-PT-RJMCMC is presented in Fig. 18. All pairs met the convergence criteria early in the sampling process, with an average length of $12.9\times10^4$. Overall, PT-RJMCMC failed once again, while AP-PT-RJMCMC performed effectively in addressing this inverse problem.

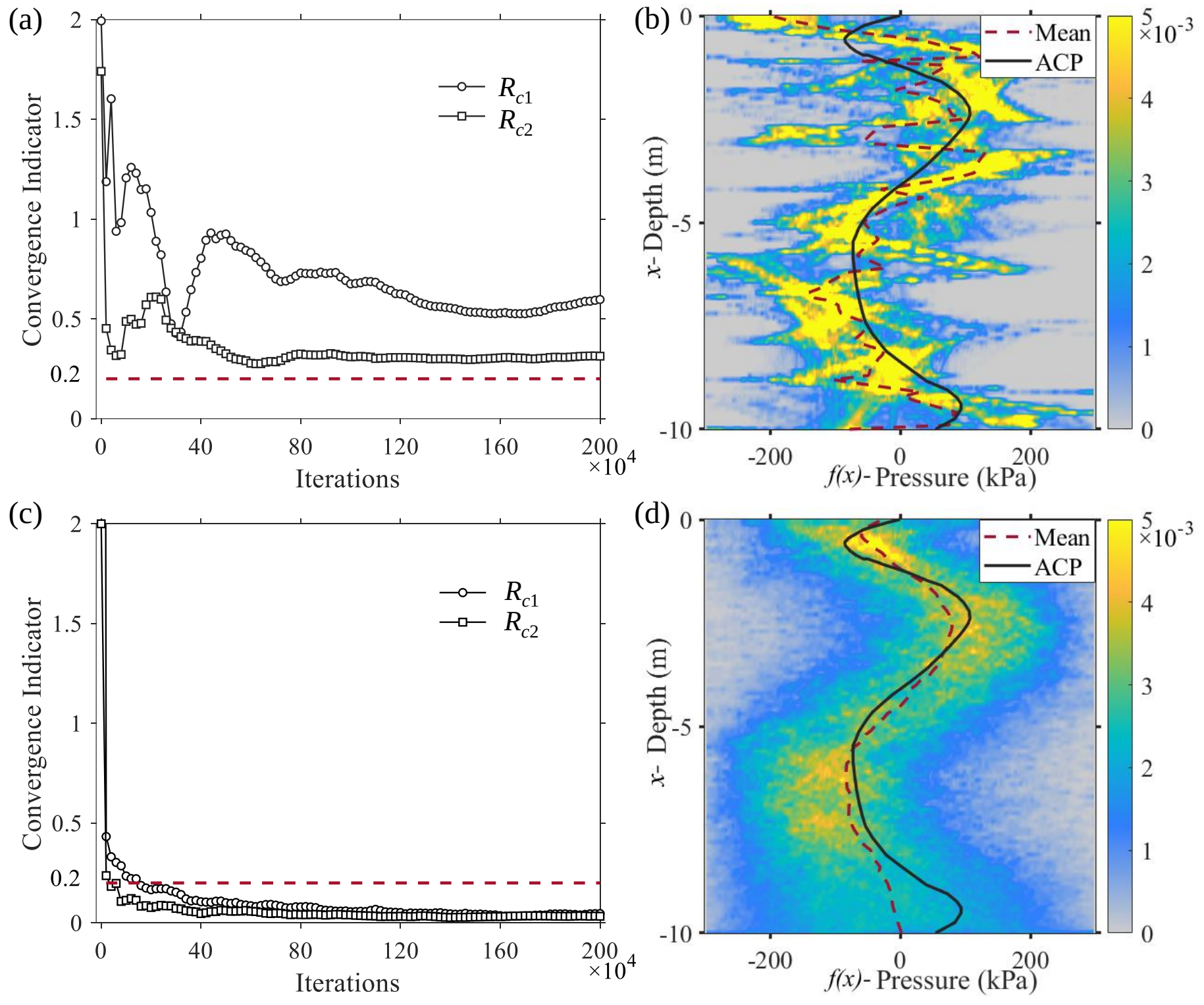

**Fig. 17.** Convergence monitoring and inversion results of the samplers in the field case: (a) convergence monitoring of a representative pair run by PT-RJMCMC; (b) inversion results by PT-RJMCMC; (c) convergence monitoring of a representative pair run by AP-PT-RJMCMC; (d) inversion results by AP-PT-RJMCMC. (ACP=actual recorded pressures)

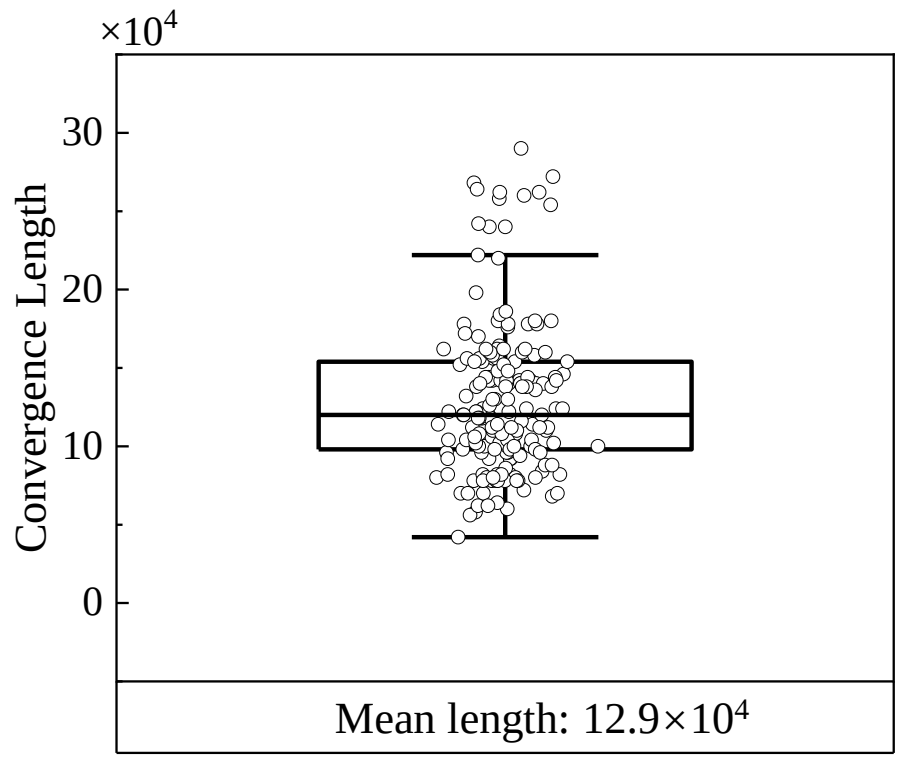

**Fig. 18**. Summary of the convergence lengths of AP-PT-RJMCMC (determined using the specified criteria involving indicators $R_{c1}$ and $R_{c2}$) for the 190 pairs in the field case (Note: PT-RJMCMC failed to meet the convergence criteria in any of the 190 pairs, resulting in the absence of data for the display of its convergence length).

## 6 Summary and Discussion

An adaptive RJMCMC algorithm is presented to address the Bayesian curve fitting problems. In comparison to conventional RJMCMC, both inter- and intra-model proposals in this algorithm can dynamically adapt based on sampling history, aiming to accelerate the sampling process. To further enhance efficiency, an auxiliary-tempered version of this algorithm is introduced by incorporating the parallel tempering (PT) technique.

Numerical examples demonstrate that the adaptive proposal (AP) algorithms are considerably more efficient than the conventional ones. Particularly noteworthy is the algorithm's performance in instances where the posterior distribution exhibits significant variability along the domain of definition. In these cases, the proposal within the adaptive algorithm (AP-PT-RJMCMC) automatically adapts to the *a priori* unknown complex posterior covariance. This feature contributes to the superior performance of the adaptive algorithm, compared to the conventional approach (PT-RJMCMC) that struggles with slower convergence.

In addition to AP, the contribution of PT should not be underestimated. A comparison between AP-PT-RJMCMC and AP-RJMCMC in the second example reveals that, despite the proposal being adapted, the adaptive algorithm may still encounter difficulties with inter-model transitions. However, PT demonstrates its ability to alleviate this limitation. Furthermore, PT facilitates early exploration of the target distribution, thereby enhancing the adaptation process.

Similar to the numerical example, the posterior distribution in practical inverse problems is typically complex and necessitates an automatically adapting proposal function. The field case study demonstrates the essential role of AP-PT-RJMCMC in achieving a stable posterior distribution, outperforming PT-RJMCMC. This suggests that this adaptive algorithm holds significant promise for addressing other complex inverse problems.

It is worth mentioning that the adaptive algorithm is essentially an extension of the Adaptive Metropolis [22] from fixed-dimensional MCMC to trans-dimensional MCMC. The key issue we addressed is exploiting the fact that the posterior covariance of the function at any set of points is well-defined and can be used to tune the proposal appropriately. In trans-dimensional problems other than curve-fitting, this same idea may not be applicable. With this addressed in this paper, it is believed further innovations (like genetic algorithms [34]) may be introduced to make the trans-dimensional sampler much more efficient in curve fitting problems.

## Acknowledgements

Z.T. was supported by Natural Science Foundation of China (Grant No. 51978523). A.L. was supported by the Engineering and Physical Sciences Research Council (EP/R034710/1). Z.T. acknowledges the invaluable support from Dr. Sam Power.

## Data Availability Statement

All data generated or used during the study are included in this paper. Code that supports the findings are available from Zhiyao.tian@bristol.ac.uk upon reasonable request, including the MATLAB script for the adaptive proposal algorithms.

## APPENDIX A: ACCEPTANCE RATIOS

This appendix presents derivation of acceptance ratios for birth, death, and move proposals. According to detailed balance, a proposal is accepted randomly with an acceptance ratio:

$$\alpha(\mathbf{m}^*|\mathbf{m}_c) = \min\{1, \frac{p(\mathbf{D}|\mathbf{m}^*)p(\mathbf{m}^*)q(\mathbf{m}_c|\mathbf{m}^*)}{p(\mathbf{D}|\mathbf{m}_c)p(\mathbf{m}_c)q(\mathbf{m}^*|\mathbf{m}_c)}|\mathbf{J}|\}, \tag{A1}$$

where $q(\mathbf{m}^*|\ \mathbf{m}_c)$ denotes the proposal function that generates a candidate model configuration $\mathbf{m}^*$ from the current model configuration $\mathbf{m}_c$, while $q(\mathbf{m}_c|\mathbf{m}^*)$ denotes the reverse proposal from $\mathbf{m}^*$ to $\mathbf{m}_c$. The terms $p(\mathbf{m})$ and $p(\mathbf{D}|\mathbf{m})$ are the prior and likelihood probabilities for a particular model configuration, respectively. $|\mathbf{J}|$ is the Jacobian matrix of the transformation from $\mathbf{m}_c$ to $\mathbf{m}^*$, which has been proven to equal 1 in algorithms where the number of knots changes by one at a step [20].

This paper focuses on cases involving uniform priors:

$$p(\mathbf{m})=p(n)p(\mathbf{r}|n)p(\mathbf{a}|n,\mathbf{r}), \tag{A2}$$

where $n$ denotes the number of knots, uniformly distributed between $n_{\max}$ and $n_{\min}$:

$$p(n)=\frac{1}{n_{\max}-n_{\min}}. \tag{A3}$$

**r** denotes location of the knots that have an equal chance to be placed on every candidate point except the endpoints:

$$p(\mathbf{r}|n)=\frac{1}{C_{N_g-2}^{n-2}}, \tag{A4}$$

where $N_g$ is number of the candidate points, and two knots are always fixed at the ends. **a** denotes the values on the knots that are independently and uniformly distributed between $a_{\max}$ and $a_{\min}$:

$$p(\mathbf{a}|n,\mathbf{r})=\prod_{i=1}^{n}\frac{1}{a_{\max}-a_{\min}}. \tag{A5}$$

The proposal function can be written as:

$$q(\mathbf{m}^*|\mathbf{m}_c) = q(n^*|n_c)q(\mathbf{r}^*|\mathbf{r}_c)q(\mathbf{a}^*|\mathbf{a}_c). \tag{A6}$$

The perturbation on dimension is always symmetric, and thus:

$$\frac{q(n^*|n_c)}{q(n_c|n^*)}=1. \tag{A7}$$

In a birth proposal, the location of this new knot is randomly chosen from the candidate points that were not occupied at the current step:

$$q(\mathbf{r}^*|\mathbf{r}_c)_{\text{birth}} = \frac{1}{C^1_{N_g-n_c}} = \frac{1}{N_g - n_c}. \tag{A8}$$

The value on the knot $a_b$ is determined by making a Gaussian perturbation on the current interpolated value $a_p$ with variance $\sigma_b^2$:

$$q(\mathbf{a}^*|\mathbf{a}_c)_{\text{birth}} = q(a_b|a_p) = \frac{1}{\sqrt{2\pi}\sigma_b}\exp[-\frac{(a_b - a_p)^2}{2\sigma_b^2}]. \tag{A9}$$

The death proposal is the reverse step of the birth proposal. A knot is deleted by choosing one randomly from the current knots except the ends, and thus,

$$q(\mathbf{r}^*|\mathbf{r}_c)_{\text{death}} = \frac{1}{C^1_{n_c-2}} = \frac{1}{n_c - 2}, \tag{A10}$$

$$q(\mathbf{a}^*|\mathbf{a}_c)_{\text{death}} = 1, \tag{A11}$$

By substituting Eq. (A2) – (A11) into Eq. (A1), the acceptance ratios for birth and death steps are obtained, respectively:

$$\alpha(\mathbf{m}^*|\mathbf{m}_c)_{\text{birth}} = \min\{1, \frac{p(\mathbf{D}|\mathbf{m}^*)}{p(\mathbf{D}|\mathbf{m}_c)} \times \frac{1}{a_{\max} - a_{\min}} \times \frac{1}{q(a_b|a_p)}\}, \tag{A12}$$

$$\alpha(\mathbf{m}^*|\mathbf{m}_c)_{\text{death}} = \min\{1, \frac{p(\mathbf{D}|\mathbf{m}^*)}{p(\mathbf{D}|\mathbf{m}_c)} \times (a_{\max} - a_{\min}) \times q(a_b|a_p)\}. \tag{A13}$$

Since the proposal for a move step is symmetric, both the proposal and prior ratios get cancelled:

$$\alpha(\mathbf{m}^*|\mathbf{m}_c)_{\text{move}} = \min\{1, \frac{p(\mathbf{D}|\mathbf{m}^*)}{p(\mathbf{D}|\mathbf{m}_c)}\}. \tag{A14}$$

## APPENDIX B: DISCUSSION ON DIMINISHING ADAPTATION

We discuss here why the diminishing adaptation condition holds for the adaptive RJMCMC algorithm. We have

$$\log(s_{c,i}) = \max\{\min\{\log(s_{c,i-1}) + \gamma_i\delta_i, \overline{s}_c\}, \underline{s}_c\}, \tag{B1}$$

where the parameters retain the same definitions as delineated in Eq. (12). We denote $\tilde{\mathbf{C}}_t = \mathbf{C}_t + \varepsilon\mathbf{I}$. Based on Eq. (10), we have

$$\tilde{\mathbf{C}}_t = \tilde{\mathbf{C}}_{t-1} + \frac{1}{t}\mathbf{A}_t, \tag{B2}$$

where $\mathbf{A}_t = \{[f_{\mathbf{m}_t}(\mathbf{z}_i) - \overline{f}_{t-1}(\mathbf{z}_i)][f_{\mathbf{m}_t}(\mathbf{z}_i) - \overline{f}_{t-1}(\mathbf{z}_i)]^{\mathrm{T}} - \mathbf{C}_{t-1}\}$ and the parameters retain the same definitions as delineated in Eq. (10). It is worth noting that $\delta_i$ and $\mathbf{A}_t$ are bounded uniformly due to acceptance probabilities being bounded and the compactness of the space, respectively. We denote by $i_t$ the number of move steps at time $t$, noting that $i_t/t \to 1/3$ almost surely due to the equal probability choice of move, birth, and death steps.

Following [Roberts and Rosenthal (2007), p. 466] [29] and in particular the arguments in [Haario et al. (2001), Proof of Theorem 1] [22], we may deduce that

$$\begin{aligned}\sup_{\mathbf{m}} \left\| P_t^{\text{move}}(\mathbf{m},\cdot) - P_{t-1}^{\text{move}}(\mathbf{m},\cdot) \right\|_{\text{TV}} &\lesssim \left\| s_{c,i_t} s_d \tilde{\mathbf{C}}_t - s_{c,i_{t-1}} s_d \tilde{\mathbf{C}}_{t-1} \right\| \\ &= s_d \left\| s_{c,i_t} \tilde{\mathbf{C}}_t - s_{c,i_{t-1}} \tilde{\mathbf{C}}_{t-1} \right\| \\ &\le s_d \left\| s_{c,i_{t-1}} \cdot \exp(\gamma_{i_t}\delta_{i_t})\{\tilde{\mathbf{C}}_{t-1} + \frac{1}{t}\mathbf{A}_t\} - s_{c,i_{t-1}} \tilde{\mathbf{C}}_{t-1} \right\| \\ &= s_d s_{c,i_{t-1}} \left\| \{\exp(\gamma_{i_t}\delta_{i_t}) - 1\}\tilde{\mathbf{C}}_{t-1} + \frac{1}{t}\exp(\gamma_{i_t}\delta_{i_t})\mathbf{A}_t \right\| \\ &\to 0 \end{aligned} \quad , \text{(B3)}$$

in probability as $t \to \infty$, since $\gamma_{i_t} \to 0$ in probability, $\delta_{i_t}$, $\tilde{\mathbf{C}}_{t-1}$, and $\mathbf{A}_t$ are bounded due to the compactness of the space, and $s_{c,i_{t-1}}$ is bounded above.

Similarly, one may deduce that the birth and death steps that depend on the estimated covariance satisfy diminishing adaptation. In particular, the relevant density to consider is the one-dimensional Gaussian PDF $q(a_b|a_p)$ given in Eq. (A9). This is Lipschitz continuous as a function of $\sigma_b^2$ uniformly in $(a_b, a_p)$ since $\sigma_b^2$ is lower bounded by construction and the space is compact. Note also that $\sigma_{b,t}^2 - \sigma_{b,t-1}^2$ is $O(1/t)$ by construction due to the update rule for $\tilde{\mathbf{C}}_t$. We now consider the kernels for a given knot $r_b$. Let $A$ be an arbitrary measurable set of states arising from a birth at knot $r_b$ from $\mathbf{m}$, and $\tilde{A} = \{a_b : \mathbf{m}^* \in A\}$. Denote by $q_t$ and $q_{t-1}$ the birth proposals at times $t$ and $t-1$ and $\alpha_t^{\text{birth}}$ and $\alpha_{t-1}^{\text{birth}}$ the respective acceptance probabilities. Then

$$\begin{aligned}&\left| P_t^{\text{birth}}(\mathbf{m},A) - P_{t-1}^{\text{birth}}(\mathbf{m},A) \right| \\ &= \left| \int_{\tilde{A}} q_t(a_b|a_p)\alpha_t^{\text{birth}}(\mathbf{m}^*|\mathbf{m}) - q_{t-1}(a_b|a_p)\alpha_{t-1}^{\text{birth}}(\mathbf{m}^*|\mathbf{m})\mathrm{d}a_b \right| \\ &\le \int_{\tilde{A}} \left| \min\{q_t(a_b|a_p), \frac{p(\mathbf{D}|\mathbf{m}^*)}{p(\mathbf{D}|\mathbf{m}_c)} \frac{1}{a_{\max} - a_{\min}}\} - \min\{q_{t-1}(a_b|a_p), \frac{p(\mathbf{D}|\mathbf{m}^*)}{p(\mathbf{D}|\mathbf{m}_c)} \frac{1}{a_{\max} - a_{\min}}\} \right| \mathrm{d}a_b \\ &\le \int_{\tilde{A}} \left| q_t(a_b|a_p) - q_{t-1}(a_b|a_p) \right| \mathrm{d}a_b \end{aligned} \quad , \text{(B4)}$$

where we have used the inequality $|a \wedge c - b \wedge c| \le |a - b|$ for $a, b, c \ge 0$, from which we may conclude that

$$\max_{\mathbf{m}} \left\| P_t^{\text{birth}}(\mathbf{m},\cdot) - P_{t-1}^{\text{birth}}(\mathbf{m},\cdot) \right\|_{\text{TV}} \to 0 \quad , \text{(B5)}$$

as $t \to \infty$ by Lipschitz continuity of $q(a_b|a_p)$ as a function of $\sigma_b^2$ and the compactness of the space.

For the death step, we have for any $\mathbf{m}^*$ resulting from the deletion of a knot

$$
\begin{aligned}
&\left|\alpha_t^{\text{death}}(\mathbf{m}^*\left|\mathbf{m}\right.) - \alpha_{t-1}^{\text{death}}(\mathbf{m}^*\left|\mathbf{m}\right.)\right| \\
&\le \frac{p(\mathbf{D}\left|\mathbf{m}^*\right.)}{p(\mathbf{D}\left|\mathbf{m}_c\right.)}(a_{\max} - a_{\min})\left|\{q_t(a_b\left|a_p\right.) - q_{t-1}(a_b\left|a_p\right.)\}\right| \\
&\to 0
\end{aligned}
\quad , \qquad \text{(B6)}
$$

as $t\to\infty$ by Lipschitz continuity of $q(a_b|a_p)$ as a function of $\sigma_b^2$, where we have used the inequality $|1 \wedge a - 1 \wedge b| \le |a - b|$ for $a, b \ge 0$. From this and the finite number of knots we may then conclude that

$$
\max_{\mathbf{m}}\left\|P_t^{\text{death}}(\mathbf{m},\cdot) - P_{t-1}^{\text{death}}(\mathbf{m},\cdot)\right\|_{\text{TV}} \to 0 \quad , \qquad \text{(B7)}
$$

as $t\to\infty$. Since the Markov chain is an adapted homogenous chain and its kernel $P_t$ is a mixture of kernels (with constant mixture weights $w_k$) that all satisfy diminishing adaptation, the triangle inequality gives

$$
\left\|\sum_k w_k P_t^{(k)}(\mathbf{m},\cdot) - \sum_k w_k P_{t-1}^{(k)}(\mathbf{m},\cdot)\right\|_{\text{TV}} \le \sum_k w_k \left\|P_t^{(k)}(\mathbf{m},\cdot) - P_{t-1}^{(k)}(\mathbf{m},\cdot)\right\| \to 0 \quad , \qquad \text{(B8)}
$$

in probability as $t\to\infty$.

## APPENDIX C: PSEUDOCODES FOR AP-PT-RJMCMC AND AP-RJMCMC

This appendix presents the pseudocodes for the adaptive proposal PT-RJMCMC and adaptive proposal RJMCMC algorithms (AP-PT-RJMCMC and AP-RJMCMC for simplicity).

Table C1. Codes for AP-PT-RJMCMC

| AP-PT-RJMCMC | |
| --- | --- |
| // Advance chains using conventional RJMCMC in the first $t_0$ steps// | |
| for $t = 1 : t_0$ | // Loop over $t_0$// |
| for $j = 1 : T$ | //Loop over $T$ chains// |
| $\mathbf{m}_{tj}$ = PT-RJMCMC ( $\mathbf{m}_{(t-1)j}$, ***Tem***($j$) ) | // conventional PT-RJMCMC// |
| **His** ( $t$, : , $j$ ) = $f_{\mathbf{m}_{tj}}(\mathbf{z}_i)$ | //Store sampling history on every candidate points// |
| end for | |
| end for | |
| // Initialize covariance matrices// | |
| for $j = 1 : T$ | //Loop over $T$ chains// |
| $\mathbf{C}_{t-1}$( : , : , $j$ ) = Cov ( **His** ( : , : , $j$ ) ) | //Calculate the initial covariance matrix// |
| end for | |
| // Advance chains using AP-RJMCMC in the remaining steps// | |
| for $t = t_0+1 : t_{\text{end}}$ | // Loop over the remaining steps// |
| for $j = 1 : T$ | //Loop over $T$ chains// |

```
        [m_tj, C_t( : , : , j )]= AP-RJMCMC ( m_(t-1)j, Tem(j), C_{t-1}( : , : , j ))      //AP-RJMCMC//
      end for
      [m_tj, Tem] = Swap (m_tj, Tem)                                    //Swap between the chains//
    end for
```

Note: AP-RJMCMC is a called function shown in Table B2. $f$ is the interpolation function (Eq. 2).

Table C2. Codes for AP- RJMCMC

AP- RJMCMC

```
// Main function of AP-RJMCMC//
function   [m_t, C_t]= AP-RJMCMC ( m_{t-1}, Tem, C_{t-1})
    u=Randsample{1,2,3}         //Randomly advance the chain with a birth, death, or move step//
    switch (u)
      case 1: [m_t, C_t]= Move ( m_{t-1}, Tem, C_{t-1})                 //Go to function Move//
      case 2: [m_t, C_t] = Birth ( m_{t-1}, Tem, C_{t-1})               //Go to function Birth//
      case 3: [m_t, C_t] = Death ( m_{t-1}, Tem, C_{t-1})               //Go to function Death//
    end switch
end function

// Function Move//
function   [m_t, C_t]= Move( m_{t-1}, Tem, C_{t-1})                     //Note: m=(n,r,a)//
    n*= n_{t-1}; r*= r_{t-1}                           //Model remains unchanged in a move step//
    j=Find_Serial_Number (r*)  //Find the index vector of the currently occupied candidate points//
    C_m=s_c s_d C_{t-1}(j, j)+s_c s_d εI      //Informed from the covariance of the sampling history//
    a*= Gaussian_Random_Number (a_{t-1}, C_m)         //Proposal with the informed covariance//
    m*=(n*,r*,a*)                                                  //Obtain a candidate sample//
    acc= Accept_Ratio_Move (m*, Tem)        //Calculate the acceptance ratio using Eq. (13)//
    s_c=Update_s_c (s_c,acc)                                      //Update s_c using Eq. (12)//
    u= Uniform_Random_Number (0, 1)                                    //Get a random number//
    if acc > u                                                          //Advance the chain//
      m_t = m*
      else
      m_t = m_{t-1}
    end if
    C_t = Recursion (C_{t-1}, f_{m_t}(z_i))   //Update the covariance matrix using Eqs. (10)–(11)//
end function

// Function Birth//
function   [m_t, C_t] = Birth( m_{t-1}, Tem, C_{t-1})                   //Note: m=(n,r,a)//
    n*= n_{t-1}+1;                                                      //Dimension plus one//
```

```
        r_birth=Randsample (Setdiff(i,r_{t-1}))        //Exclude the occupied points and randomly sample one//
        r*= [r_birth, r_{t-1}]                           //Get a candidate vector r*//
        j=Find_Serial_Number (r_birth)                  //Find the index number of the newly created point//
        σ_b^2=s_d C_{t-1}(j, j)+s_d εI                  // Informed from the variance of the sampling history //
        a_birth=Gaussian_Random_Number (f_{m_{t-1}}(r_birth), σ_b^2)   //Proposal with the informed variance//
        a*= [a_birth, a_{t-1}]
        m*=(n*,r*,a*)                                    //Obtain a candidate sample//
        acc= Accept_Ratio_Birth (m*, Tem, σ_b^2)         //Calculate the acceptance ratio with Eq. (13)//
        u= Uniform_Random_Number (0, 1)                  //Get a random number//
        if acc > u                                       //Advance the chain//
            m_t = m*
            else
            m_t = m_{t-1}
        end if
        C_t = Recursion (C_{t-1}, f_{m_t}(z_i))          // Update the covariance matrix using Eqs. (10)–(11)//
end function

// Function Death//
function   [m_t, C_t]= Death( m_{t-1}, Tem, C_{t-1})    //Note: m=(n,r,a)//
        n*= n_{t-1} – 1;                                 //Dimension minus one//
        r_death=Randsample (r_{t-1})                     //Randomly sample a candidate knot to delete//
        r*= Setdiff(r_{t-1}, r_death)                    //Delete this knot//
        a*= Setdiff(a_{t-1}, a_{r_death})                //Delete this knot//
        m*=(n*,r*,a*)                                    //Get the candidate sample//
        j=Find_Serial_Number (r_death)                  //Find the index number of the deleted knot//
        σ_b^2=s_d C_{t-1}(j, j)+s_d εI                  // Informed from the variance of the sampling history //
        acc= Accept_Ratio_Death (m*, Tem, σ_b^2)         //Calculate the acceptance ratio with Eq. (13)//
        u= Uniform_Random_Number (0, 1)                  //Get a random number//
        if acc > u                                       //Advance the chain//
            m_t = m*
            else
            m_t = m_{t-1}
        end if
        C_t = Recursion (C_{t-1}, f_{m_t}(z_i))          // Update the covariance matrix using Eqs. (10)–(11)//
end function
```

Note: when AP-RJMCMC is not in a PT framework, *Tem* should be set as 1.

## APPENDIX D: THE FORWARD MODEL FOR THE FIELD EXAMPLE

This appendix presents the forward model, $d_i=g(f, x_i)$, to predict the deformation $d_i$ on a concrete pile at location $x_i$ under any given pressures $f$. The concrete pile is conceptualized as an elastic beam, with its mechanical behavior described by a fourth-order partial differential equation:

$$EI(x)\frac{\partial^4 d(x)}{\partial x^4} = f(x), \tag{D1}$$

where $x$ represents the coordinate along the pile; $d(x)$ is the function that describes the structural deformations along $x$; $EI$ is the flexural rigidity of the beam that may vary along $x$. The Finite Element Method (FEM) is commonly employed to solve Eq. (D1). More details on FEM can be found in Griffiths [35]. Specifically, Eq. (D1) is discretized into

$$\mathbf{d} = \mathbf{K}^{-1}\mathbf{f}, \tag{D2}$$

where, the pile is discretized into numerous small elements with a length of $L$ (101 elements in this example); **d** denotes the deformation vector at the nodes of the elements; **f** denotes the nodal force vector on the elements that is equivalent to $f(x)$ following the transformation rules of virtual work [36]; **K** is the global stiffness matrix that is assembled by the element stiffness matrix $\mathbf{k}^e$ using the criterion outlined in Huebner et al. [37]. Let $x_e$ to be the location of element $e$, and $\xi = x - x_e$ , then

$$\mathbf{k}^e = \int_{x_e}^{x_e+L} EI(x)\frac{d^2 N_i}{dx^2}\frac{d^2 N_j}{dx^2}dx, \tag{D3}$$

where,

$$\begin{aligned} N_1 &= \frac{2}{L^3}\xi^3 - \frac{3}{L^2}\xi^2 + 1, \quad N_2 = \frac{1}{L^2}\xi^3 - \frac{2}{L}\xi^2 + \xi \\ N_3 &= -\frac{2}{L^3}\xi^3 + \frac{3}{L^2}\xi^2, \qquad N_4 = \frac{1}{L^2}\xi^3 - \frac{1}{L}\xi^2 \end{aligned}. \tag{D4}$$

**f** is also assembled by the nodal forces $\mathbf{f}^e$, where

$$\mathbf{f}^e = \begin{bmatrix} 1 & 0 & -\frac{3}{L^2} & \frac{2}{L^3} \\ 0 & 1 & -\frac{2}{L} & \frac{1}{L^2} \\ 0 & 0 & \frac{3}{L^2} & -\frac{2}{L^2} \\ 0 & 0 & -\frac{1}{L} & \frac{1}{L^2} \end{bmatrix} \begin{Bmatrix} F_{p0} \\ F_{p1} \\ F_{p2} \\ F_{p3} \end{Bmatrix}, \tag{D5}$$

where,

$$F_{p0} = \int_{z_e}^{z_e+L} f(x)dx \qquad F_{p1} = \int_{z_e}^{z_e+L} f(x)\xi dx$$
$$F_{p2} = \int_{z_e}^{z_e+L} f(x)\xi^2 dx \quad F_{p3} = \int_{z_e}^{z_e+L} f(x)\xi^3 dx, \qquad \text{(D6).}$$